\documentclass[a4paper,11pt]{revtex4}
\usepackage{graphicx}  % standard LaTeX graphics tool;
\usepackage{amsmath}   % For getting proper math eqns
\usepackage{amssymb}   % Used for getting various symbols eg. gtrsim, lesssim etc.
\usepackage{bm} % For bold math (esp of lower greek letters)
\usepackage{dcolumn}% Align table columns on decimal point
\usepackage{color}
\usepackage{mathrsfs}
\usepackage{amsfonts}
\usepackage{varioref}
\usepackage{mathrsfs}
\usepackage{graphicx}
\usepackage{latexsym}
\usepackage{amsmath}
\usepackage{amssymb}
\usepackage{textcomp}
\usepackage{amsbsy}
\usepackage{graphics}
\usepackage{epstopdf}
\usepackage{color}
\usepackage[caption=false]{subfig}

\RequirePackage[colorlinks,citecolor=blue,urlcolor=magenta,linkcolor=blue]{hyperref}
\input epsf

\allowdisplaybreaks[4]

\begin{document}

\tolerance=5000

\title{Primordial gravitational waves in horizon cosmology and constraints on entropic parameters}

\author{Sergei~D.~Odintsov$^{1,2}$\,\thanks{odintsov@ieec.uab.es},
Simone~D'Onofrio$^{2}$\,\thanks{donofrio@ice.csic.es},
Tanmoy~Paul$^{3}$\,\thanks{tanmoy.paul@visva-bharati.ac.in}} \affiliation{
$^{1)}$ ICREA, Passeig Luis Companys, 23, 08010 Barcelona, Spain\\
$^{2)}$ Institute of Space Sciences (ICE, CSIC) C. Can Magrans s/n, 08193 Barcelona, Spain\\
$^{3)}$ Department of Physics, Visva-Bharati University, Santiniketan 731235}

%\date{}

\tolerance=5000

\begin{abstract}
The hallmark of the 4-parameter generalized entropy is that it can represent various known entropies proposed so far for suitable limits of the parameters, and thus the 4-parameter generalized entropy can be applied on additive as well as on non-additive systems. Thereafter the proposal of the generalized entropy, it becomes important to constrain the corresponding parameters. In the present work, we intend to do this from the perspective of primordial gravitational waves (GWs) generated during inflation. In the background level, the generalized entropy successfully drives a viable and smooth evolution of the universe, particularly from inflation to reheating followed by a radiation era, for certain ranges of the entropic parameters. Consequently we investigate whether such viable ranges of the entropic parameters (coming from the background level) allow the primordial GWs spectrum to pass through the sensitivity curves of various GWs observatories. Therefore, if the future observatories can detect the signal of primordial GWs, then our theoretical expectation carried in the present work may provide a possible tool for the measurement of the generalized entropic parameters.
\end{abstract}
%%%%%%%%%%%%%%%%%%%%%%%%%%%%%%%%%%%%%%%%%%%%%%%%%%%%%%%%%%%%%%%%%%%%%%%%%%%%%%%%%%%%%%%%%%%%%%%%%%%
%%%%%%%%%%%%%%%%%%%%%%%%%%%%%%%%%%%%%%%%%%%%%%%%%%%%%%%%%%%%%%%%%%%%%%%%%%%%%%%%%%%%%%%%%%%%%%%%%%%
%%%%%%%%%%%%%%%%%%%%%%%%%%%%%%%%%%%%%%%%%%%%%%%%%%%%%%%%%%%%%%%%%%%%%%%%%%%%%%%%%%%%%%%%%%%%%%%%%%%
%\newpage
%%%%%%%%%%%%%%%%%%%%%%%%%%%%%%%%%%%%%%%%%%%%%%%%%%%%%%%%%%%%%%%%%%%%%%%%%%%%%%%%%%%%%%%%%%%%%%%%%%%
%%%%%%%%%%%%%%%%%%%%%%%%%%%%%%%%%%%%%%%%%%%%%%%%%%%%%%%%%%%%%%%%%%%%%%%%%%%%%%%%%%%%%%%%%%%%%%%%%%%
%%%%%%%%%%%%%%%%%%%%%%%%%%%%%%%%%%%%%%%%%%%%%%%%%%%%%%%%%%%%%%%%%%%%%%%%%%%%%%%%%%%%%%%%%%%%%%%%%%%
%\pacs{}

\maketitle

\section{Introduction}

Bekenstein-Hawking entropy of black hole, for the first time, connects the laws of thermodynamics with spacetime gravity \cite{Bekenstein:1973ur, Hawking:1975vcx}. The interesting nature of Bekenstein-Hawking entropy is that it depends on the area of the event horizon of the black hole, unlike to the usual thermodynamic system where the entropy scales by the volume of the system under consideration \cite{Bardeen:1973gs, Wald:1999vt, Jacobson:1995ab}. This is important to understand the microscopic root of black hole entropy, which however, is still questionable and requires proper justification. Based on the distinctive nature of Bekenstein-Hawking entropy, some other forms of black hole entropy were proposed, like the Tsallis entropy and the R\'{e}nyi entropy depending on the non-additive statistics \cite{Tsallis:1987eu,Renyi}. Thereafter the Barrow entropy has been proposed, which may encode the quantum nature of black hole through the fractal structures \cite{Barrow:2020tzx}. Some other well known entropies are Sharma-Mittal entropy \cite{SayahianJahromi:2018irq}, Kaniadakis entropy \cite{Kaniadakis:2005zk}, or even, the Loop Quantum Gravity entropy \cite{Liu:2021dvj} etc.

The growing interest of horizon thermodynamics is also extended to the sector of cosmology where the laws of thermodynamics is generally applied to the apparent horizon having radius $R_\mathrm{h} = 1/H$ (with $H$ being the Hubble parameter at an instant) \cite{Cai:2005ra, Akbar:2006kj, Cai:2006rs, Akbar:2006er,
Paranjape:2006ca, Jamil:2009eb, Cai:2009ph, Wang:2009zv, Jamil:2010di, Gim:2014nba, DAgostino:2019wko, Sanchez:2022xfh, Cognola:2005de,Nojiri:2022nmu,Nojiri:2023wzz,Odintsov:2024hzu}. Actually being a null surface, the apparent horizon divides the observable universe from an unobservable one. Now due to the cosmic evolution, there exists a flux of the normal matter fields from inside to outside of the horizon. This flux causes a decrease of the matter fields' entropy inside of the horizon, which in turn demands an entropy of gravitational field in order to validate the second law of thermodynamics for the combined system of ``matter field $+$ gravitational field'' \cite{Odintsov:2024hzu}. Such gravitational entropy is accounted through the entropy of the apparent horizon. At this stage it deserves mentioning that similar to the case of black hole, the microscopic origin of entropy of the cosmic apparent horizon still demands a proper understanding (one may see \cite{Nojiri:2023bom} for recent developments in this regard). Based on the thermodynamics of the apparent horizon, the Bekenstein-Hawking like entropy leads to the usual Friedmann equations with normal matter fields. However if the apparent horizon is considered to have a different form of entropy (compared to the Bekenstein-Hawking like), then it results to a modified cosmological scenario and the modification appears through an effective entropic energy density. Such modified cosmic scenario turns out to have rich consequences in various cosmological directions started from inflation (or bounce) to dark energy era \cite{Cai:2005ra, Akbar:2006kj, Cai:2006rs, Akbar:2006er,
Paranjape:2006ca, Jamil:2009eb, Cai:2009ph, Wang:2009zv, Jamil:2010di, Gim:2014nba, DAgostino:2019wko, Sanchez:2022xfh, Cognola:2005de,Nojiri:2022nmu,Nojiri:2023wzz,Odintsov:2024hzu,Barrow:2020kug,Nojiri:2019skr,Jizba:2023fkp,Jizba:2024klq,Nojiri:2022aof,Nojiri:2022dkr,Odintsov:2022qnn}. Some variants of Bekenstein-Hawking entropy that are frequently used in cosmology are the Tsallis entropy, the R\'{e}nyi entropy, the Barrow entropy, the Sharma-Mittal entropy, the Kaniadakis entropy etc. The interesting fact is that all of these entropies share some common properties like --- they are monotonically increasing functions of the Bekenstein-Hawking entropy variable ($S$), and moreover, they all vanish at the limit $S \rightarrow 0$. Owing to such common properties, the question that may naturally arise --- `` does there exist a generalized form of entropy that can bring all the known entropies aforementioned within a single umbrella ?'' In spirit of this, generalized form of entropy has been recently proposed in \cite{Nojiri:2022aof,Nojiri:2022dkr,Odintsov:2022qnn}, containing few parameters and is able to represent all the known entropies proposed so far for suitable limits of the parameters. In particular, two different generalized entropies have been proposed having 6-parameter and 4-parameter respectively \cite{Nojiri:2022aof,Nojiri:2022dkr}, however the conjecture in \cite{Nojiri:2022dkr} states that the minimum number of parameters required in a generalized entropy function that can generalize all the aforementioned entropies is equal to four. Thereby the minimal version of generalized entropy is the 4-parameter generalized entropy given by,
\begin{eqnarray}
 S_\mathrm{g}\left[\alpha_+,\alpha_-,\beta,\gamma \right]&=&\frac{1}{\gamma}\left[\left(1 + \frac{\alpha_+}{\beta}~S\right)^{\beta}
 - \left(1 + \frac{\alpha_-}{\beta}~S\right)^{-\beta}\right]~~,
 \label{intro-1}
\end{eqnarray}
on which, we will concentrate in the present work. Here $\alpha_{\pm}, \beta, \gamma$ are the parameters assumed to be positive in order to make $S_\mathrm{g}$ a monotonic increasing function w.r.t. the Bekenstein-Hawking entropy ($S$). The cosmological implications of generalized entropy(ies) are discussed in \cite{Nojiri:2022dkr,Odintsov:2022qnn,Odintsov:2023vpj,Bolotin:2023wiw,Lymperis:2023prf}.

Thereafter the proposal of 4-parameter generalized entropy, it becomes important to constrain the corresponding parameters. In the present work, we are motivated to do this from the perspective of primordial gravitational waves (GWs). Primordial GWs carry the footprints of universe during its early stages. With a variety of GW detectors being proposed to operate over a wide range of frequencies, there is great expectation that observations of primordial GWs can provide us with an unprecedented window to the physics operating during inflation and reheating, including the underlying physics of primordial black holes \cite{Turner:1993vb,Boyle:2005se,Nakayama:2008wy,Alabidi:2012ex,Haque:2021dha,Maiti:2024nhv,Sasaki:2018dmp,Sasaki:2016jop,Domenech:2020ssp}. Some proposed GWs observatories, covering a wide range of frequencies, include advanced LIGO \cite{LIGOScientific:2016jlg}, ET \cite{Punturo:2010zz}, DECIGO \cite{Seto:2001qf}, LISA \cite{Amaro-Seoane:2012aqc}, BBO \cite{Crowder:2005nr}, SKA \cite{Janssen:2014dka}. For the purpose of our work, we first need to investigate whether the 4-parameter generalized entropy can successfully trigger a viable and smooth evolution of the universe, particularly from inflation to reheating followed by a radiation dominated era. Consequently we will turn to study the primordial GWs in the present context of horizon cosmology. In particular, we will examine whether the viable ranges of the entropic parameters (coming from the background level) can enhance the tensor perturbation amplitude and allow the primordial GWs spectrum today to pass through the sensitivity curves of various GWs observatories like LISA, DECIGO, BBO etc. This in turn may provide a possible tool for the measurement of the generalized entropic parameters. In the present work, we will consider classical entropic cosmology but it would be very interesting to take into account also quantum effects, in the way \cite{Glavan:2024elz,Glavan:2023tet}.

Before we proceed further, a few clarifications concerning the conventions and notations that we shall adopt are in order. We will work in natural units with $\hbar = c=1$, and we will adopt the metric signature given by $\left(-,+,+,+\right)$. A suffix 'f' with some quantity represents the same at the end of inflation, while a suffix 're' denotes the end of reheating. For instance, if $a$ is the scale factor then $a_\mathrm{f}$ and $a_\mathrm{re}$ represent the scale factor at the of inflation and at the end of reheating, respectively. Moreover an overdot and an overprime designates the differentiation w.r.t the cosmic time ($t$) and the conformal time ($\eta$) respectively. We would also like to mention that the frequency ($f$) of tensor perturbation is connected to the corresponding wave number ($k$) via:
\begin{eqnarray}
 f = 1.55\times10^{-15}\left(\frac{k}{1~\mathrm{Mpc}^{-1}}\right)\mathrm{Hz}.
 \label{intro-2}
\end{eqnarray}

\section{Thermodynamics of apparent horizon and cosmological field equations from the 4-parameter generalized entropy}\label{SecII}

We consider a spatially flat Friedmann-Lema\^{i}tre-Robertson-Walker (FLRW) universe with the following line element:
\begin{align}
\label{dS7}
ds^2 = - dt ^2 + a( t )^2 \left( d r ^2 + r ^2 {d\Omega}^2 \right) \, ,
\end{align}
where ${d\Omega}^2$ is the line element of 3 dimensional sphere of unit radius (particularly on the surface of the sphere usually designated by the coordinates $\theta$ and $\phi$). We also define the line element perpendicular to the sphere as,
\begin{align}
\label{dS7B}
d{s_\perp}^2 = \sum_{M,N=0,1} h_{\mu\nu} dx^M dx^N = - dt ^2 + a( t )^2 d r ^2 \, .
\end{align}
The radius of the apparent horizon $R_\mathrm{h}=R\equiv a(t)r$ for the FLRW universe is given by the solution of the equation
$h^{MN} \partial_M R \partial_N R = 0$ (see \cite{Cai:2005ra, Akbar:2006kj, Sanchez:2022xfh}) which immediately leads to,
\begin{align}
\label{dS14A}
R_\mathrm{h}=\frac{1}{H}\, ,
\end{align}
for the metric (\ref{dS7}), with $H\equiv \frac{1}{a}\frac{da}{d t }$ represents the Hubble parameter of the universe. It may be noted that the apparent horizon in the case of a spatially flat FLRW universe becomes equal to the Hubble radius. The surface gravity $\kappa$ on the apparent horizon is defined as \cite{Cai:2005ra}
\begin{align}
\label{SG3}
\kappa= \left. \frac{1}{2\sqrt{-h}} \partial_M \left( \sqrt{-h} h^{MN} \partial_N R \right) \right|_{R=R_\mathrm{h}}\, .
\end{align}
For the metric of Eq.~(\ref{dS7}), we have $R=a r $ and obtain
\begin{align}
\label{SG2}
\kappa = - \frac{1}{R_\mathrm{h}} \left\{ 1 + \dot{H}\left(\frac{{R_\mathrm{h}}^2}{2}\right) \right\} \, ,
\end{align}
where the following expression is used,
\begin{align}
\label{dS14AB}
\dot R_\mathrm{h} = - H\dot{H} {R_\mathrm{h}}^3 \, .
\end{align}
The surface gravity of Eq.~(\ref{SG2}) is related with the temperature via $T_\mathrm{h} = \kappa/(2\pi)$, i.e.,
\begin{align}
\label{AH2}
T_\mathrm{h} \equiv \frac{\left| \kappa \right|}{2\pi}
= \frac{1}{2\pi R_\mathrm{h}} \left| 1 - \frac{\dot R_\mathrm{h}}{2 H R_\mathrm{h}} \right|
= \frac{H}{2\pi} \left| 1 + \frac{\dot{H}}{2H^2} \right|\, ,
\end{align}
in terms of the Hubble parameter and its derivative.
Consequently, we may associate an entropy ($S_\mathrm{h}$) to the apparent horizon, which in turn follows the thermodynamic law given by \cite{Akbar:2006kj, Sanchez:2022xfh,Nojiri:2023wzz,Odintsov:2024hzu},
\begin{align}
TdS_\mathrm{h} = -dE + WdV\, ,
\label{law-1}
\end{align}
where $V = \frac{4\pi}{3}R_\mathrm{h}^3$ is the volume of the space enclosed by the apparent horizon. Moreover, $E = \rho V$ is the total internal energy of the matter fields inside of the horizon, and $W = \frac{1}{2}\left(\rho - p\right)$ denotes the work density by the matter fields \cite{Akbar:2006kj, Sanchez:2022xfh}. The factor $\frac{1}{2}\left(\rho - p\right)$ arises in the expression of $W$ due to the fact that the work done is accounted by the trace of the energy-momentum tensor of the matter fields along the perpendicular direction of the apparent horizon. Eq.~(\ref{law-1}) clearly depicts that the entropy of the apparent horizon generates from the two terms given by $-dE$ and $WdV$, respectively. Actually both of these terms indicate a flux of the matter fields from inside to the outside of the horizon leading to a decrease of the matter fields' entropy. This in turn demands an entropy of gravitational field in order to validate the second law of thermodynamics for the combined system of ``matter field $+$ gravitational field'' \cite{Odintsov:2024hzu}. Such gravitational entropy is reflected through the horizon entropy.

Now if the entropy of the apparent horizon is considered to have a form of the 4-parameter generalized entropy (see Eq.~(\ref{intro-1})), then Eq.~(\ref{law-1}) along with the conservation of matter fields give rise to the following field equations:
\begin{align}
\frac{1}{\gamma}\left[\alpha_{+}\left(1 + \frac{\pi \alpha_+}{\beta GH^2}\right)^{\beta - 1}
+ \alpha_-\left(1 + \frac{\pi \alpha_-}{\beta GH^2}\right)^{-\beta-1}\right]\dot{H} = -4\pi G\left(\rho + p\right)~~,
\label{FRW-1}
\end{align}
and
\begin{align}
\frac{GH^4\beta}{\pi\gamma}&\,\left[ \frac{1}{\left(2+\beta\right)}\left(\frac{GH^2\beta}{\pi\alpha_-}\right)^{\beta}~
2F_{1}\left(1+\beta, 2+\beta, 3+\beta, -\frac{GH^2\beta}{\pi\alpha_-}\right) \right. \nonumber\\
&\, \left. + \frac{1}{\left(2-\beta\right)}\left(\frac{GH^2\beta}{\pi\alpha_+}
\right)^{-\beta}~2F_{1}\left(1-\beta, 2-\beta, 3-\beta, -\frac{GH^2\beta}{\pi\alpha_+}\right) \right] = \frac{8\pi G\rho}{3} + \frac{\Lambda}{3} \,,
\label{FRW-2}
\end{align}
respectively, where $2F_1(a,b,c,x)$ denotes the Hypergeometric function, $\rho$ (and $p$) symbolize the energy density (and pressure) of the normal matter fields inside the horizon and $\Lambda$ is the integration constant (also known as the cosmological constant). Thus as a whole, Eq.~(\ref{FRW-1}) and Eq.~(\ref{FRW-2}) represent the modified Friedmann equations corresponding to the 4-parameter generalized entropy. Such modified field equations may give rise to some interesting cosmology which, in turn, will put viable constraints on the entropic parameters ($\alpha_{\pm},\beta, \gamma$). As mentioned in the introduction, we intend to scan such viable ranges of the entropic parameters from the perspective of primordial GWs.

\section{Background evolution from the 4-parameter generalized entropy}

Before proceeding to the primordial GWs, we first need to go through the background evolution. In particular, we will examine whether the horizon cosmology corresponding to the 4-parameter generalized entropy ($S_\mathrm{g}$) leads to a viable and smooth evolution of the universe, started from inflation to reheating followed by a radiation era.

\subsection{A successful inflation}\label{sec-inf}

During the early stage of the universe, we consider $\rho = p = \Lambda = 0$ (as the normal matter gets diluted due to the fast expansion of the universe during its early stage), due to which, Eq.~(\ref{FRW-1}) becomes,
\begin{align}
\frac{1}{\gamma}\left[\alpha_{+}\left(1 + \frac{\pi \alpha_+}{\beta GH^2}\right)^{\beta - 1}
+ \alpha_-\left(1 + \frac{\pi \alpha_-}{\beta GH^2}\right)^{-\beta-1}\right]\dot{H} = 0~~.
\label{FRW-2-inf}
\end{align}
Here we would like to mention that the normal matters, if present at early universe, get diluted due to the fast expansion of the universe. However later we will show that during the reheating epoch, the energy density corresponding to $S_\mathrm{g}$ gets transformed to normal matter energy density that connects the reheating era to the Standard Big-Bang Cosmology (SBBC). Coming back to Eq.~(\ref{FRW-2-inf}), it admits a $constant~Hubble~parameter$ as the solution which depicts that the entropic cosmology based on the generalized entropy $S_\mathrm{g}$ can lead to a de-Sitter (dS) inflation during the early universe. However a dS inflation is plagued with some problems like it has no exit, or, the primordial curvature perturbation for a dS inflation becomes exactly flat which is not consistent with the Planck data. Thus in order to describe a quasi-dS inflationary scenario, here we will consider that the entropic parameters are not strictly constant, rather they slightly vary with the cosmic expansion of the universe. The entropic cosmology with varying exponents are well studied in \cite{Nojiri:2019skr,Odintsov:2023vpj}. In particular, here we consider the parameter
$\gamma$ to vary by the following way and the other parameters (i.e., $\alpha_+$, $\alpha_-$ and $\beta$) remain constant with $t$:
\begin{align}
\gamma(N) = \mathrm{exp}\left[\int_{N_\mathrm{f}}^{N}\sigma(N) dN\right]~~.
\label{gamma function}
\end{align}
Here $N$ denotes the e-folding number with $N_\mathrm{f}$ being the total e-folding number of the inflationary era.

During inflation, $\sigma(N)$ is of the form,
\begin{align}
\sigma(N) = \sigma_0 + \mathrm{e}^{-\left(N_\mathrm{f} - N\right)}~~;~~~~\mathrm{during~inflation} \,,
\label{sigma function}
\end{align}
where $\sigma_0$ is a constant. The term $\mathrm{e}^{-\left(N_\mathrm{f} - N\right)}$ in the above expression becomes effective only around $N = N_\mathrm{f}$, i.e near the end of inflation, and thus the term proves to be useful to make an exit of the inflation under consideration.
%Using Eq.(\ref{sigma function}), we give a plot of $\sigma(N)$ vs. $N$, see Fig.[\ref{plot-sigma}] where we take $N_\mathrm{f} = 55$ and $\sigma_0 = 0.015$ which actually lies within the viable parameteric regime in respect to the Planck data (as we will show later in this section).

%\begin{figure}[!h]
%\begin{center}
%\centering
%\includegraphics[width=3.5in,height=2.5in]{sigma.pdf}
%\caption{$\sigma(N)$ vs. $N$ for $\sigma_0 = 0.015$ and $N_\mathrm{f} = 55$.}
% \label{plot-sigma}
%\end{center}
%\end{figure}

%Fig.[\ref{plot-sigma}] clearly demonstrates that $\sigma(N)$ remains almost constant at $\sigma = \sigma_0$ during most of the inflationary e-fold, however it starts to vary around the end of inflation and reaches at $\sigma = 1+\sigma_0$ at $N = N_\mathrm{f}$.
In such scenario where $\gamma$ varies with $N$, the Friedmann equation turns out to be \cite{Odintsov:2023vpj},
\begin{align}
 -\left(\frac{2\pi}{G}\right)
\left[\frac{\alpha_+\left(1 + \frac{\alpha_+}{\beta}~S\right)^{\beta-1} + \alpha_-\left(1 + \frac{\alpha_-}{\beta}~S\right)^{-\beta-1}}
{\left(1 + \frac{\alpha_+}{\beta}~S\right)^{\beta} - \left(1 + \frac{\alpha_-}{\beta}~S\right)^{-\beta}}\right]\frac{1}{H^3}\frac{dH}{dN} = \sigma(N) \,,
\label{FRW-eq-viable-inf-main1}
\end{align}
on solving which for the Hubble parameter $H = H(N)$, we get,
\begin{align}
H(N) = 4\pi M_\mathrm{Pl}\sqrt{\frac{\alpha_+}{\beta}}
\left[\frac{2^{1/(2\beta)}\exp{\left[-\frac{1}{2\beta}\int^{N}\sigma(N)dN\right]}}
{\left\{1 + \sqrt{1 + 4\left(\alpha_+/\alpha_-\right)^{\beta}\exp{\left[-2\int^{N}\sigma(N)dN\right]}}\right\}^{1/(2\beta)}}\right] \,.
\label{solution-viable-inf-2}
\end{align}
During inflation, $\sigma(N)$ is of the form of Eq.(\ref{sigma function}) and consequently we obtain $\int_0^{N}\sigma(N)dN = N\sigma_0 + \mathrm{e}^{-(N_\mathrm{f} - N)} - \mathrm{e}^{-N_\mathrm{f}}$. The above solution of $H(N)$ immediately leads to the slow roll parameter $\epsilon = -d\mathrm{ln}H/dN$ as follows:
\begin{align}
\epsilon(N) = \frac{\sigma(N)}
{2\beta\sqrt{1 + 4\left(\alpha_+/\alpha_-\right)^{\beta}\exp{\left[-2\int_0^{N}\sigma(N)dN\right]}}} \,.
\label{slow roll parameter}
\end{align}
Eq.~(\ref{slow roll parameter}) clearly reveals that for $\sigma(N) = 0$ (or equivalently, $\gamma(N)=\mathrm{constant}$), the slow roll parameter vanishes resulting to a dS inflation that we have got in the previous part. However with a varying $\gamma(N)$, $\epsilon(N)$ does not vanish which in turn may lead to a quasi-dS inflation. Due to $\sigma(N) > 0$, the $\epsilon(N)$ from Eq.~(\ref{slow roll parameter}) becomes an increasing function with $N$; as a result, we may fix the entropic parameters in such a way that $\epsilon(N)$ remains less than unity during $N < N_\mathrm{f}$ and reaches to unity at $N = N_\mathrm{f}$ which results to the end of inflation. Clearly the condition $\epsilon(N_\mathrm{f}) = 1$ immediately leads to the following relation between the entropic parameters from Eq.(\ref{slow roll parameter}) as follows:
\begin{align}
\beta = \frac{(1 + \sigma_0)}
{2\sqrt{1 + 4\left(\alpha_+/\alpha_-\right)^{\beta}\exp{\left[-2\left(1 + \sigma_0N_\mathrm{f}\right)\right]}}} \,,
\label{end of inflation}
\end{align}
where we use $\int_0^{N_\mathrm{f}}\sigma(N)dN = 1+\sigma_0N_\mathrm{f}$. The inflationary observables, in particular, the spectral tilt for the primordial curvature perturbation ($n_s$) and the tensor-to-scalar ratio ($r$) are defined by \cite{Schwarz:2001vv}:
\begin{eqnarray}
 n_s = 1- 2\epsilon - d\mathrm{ln}\epsilon/dN~~~~~~~~~~~\mathrm{and}~~~~~~~~~~~r=16\epsilon \,,
 \label{new-ns-r}
\end{eqnarray}
respectively (both of these observables are defined at the horizon crossing instance of the CMB scale mode $\sim 0.05\mathrm{Mpc}^{-1}$, which is considered to be $N=0$ in the present context). Here it may be mentioned that these expressions of $n_s$ and $r$ considered in the present context are similar to that of a slow roll inflation in canonical scalar-tensor (ST) theory. This is because the fact that the solution of $H(N)$ of Eq.~(\ref{solution-viable-inf-2}) can mimic slow roll inflation under a canonical ST theory with a suitable scalar potential along with a certain evolution of the scalar field (see Appendix Sec.~[\ref{sec-appendix}]).

By using Eq.~(\ref{slow roll parameter}), the $n_s$ and $r$ from the above Eq.~(\ref{new-ns-r}) are evaluated as follows:
\begin{align}
n_s = 1 - \frac{\sigma_0}{\beta\sqrt{1 + \mathrm{exp}\left[2\left(1 + \sigma_0N_\mathrm{f}\right)\right]\left[\left(\frac{1+\sigma_0}{2\beta}\right)^2 - 1\right]}} - \frac{\sigma_0\left[\left(\frac{1+\sigma_0}{2\beta}\right)^2 - 1\right]}{\mathrm{exp}\left[-2\left(1 + \sigma_0N_\mathrm{f}\right)\right] + \left[\left(\frac{1+\sigma_0}{2\beta}\right)^2 - 1\right]} \,,
\nonumber
\end{align}
and 
\begin{align}
r = \frac{8\sigma_0}{\beta\sqrt{1 + \mathrm{exp}\left[2\left(1 + \sigma_0N_\mathrm{f}\right)\right]\left[\left(\frac{1+\sigma_0}{2\beta}\right)^2 - 1\right]}}
\label{r final form}
\end{align}
respectively. To obtain such expressions of $n_s$ and $r$, we use Eq.(\ref{end of inflation}), i.e the above forms of $n_s$ and $r$ contain the information of $\epsilon(N_\mathrm{f}) = 1$. Confronting the theoretical expectations of $n_s$ and $r$ with the Planck data ($n_s = 0.9649 \pm 0.0042$ and $r < 0.064$) \cite{Planck:2018jri}, we get the following viable ranges of the entropic parameters (see Table.~[\ref{Table-1}]):

\begin{table}[h]
  \centering
 {%
  \begin{tabular}{|c|c|c|c|}
   \hline
    Viable choices of $N_\mathrm{f}$ & Range of $\beta$ & Range of $\sigma_0$ & Range of $\left(\frac{\alpha_+}{\alpha_-}\right)^{\beta}$\\

   \hline
   (1) Set-1: $N_\mathrm{f} = 50$ & $\beta= (0,0.35]$ & $\sigma_0=[0.0132,0.0170]$ & $\left(\alpha_+/\alpha_-\right)^{\beta} \geq 7.5$\\
   \hline
    (2) Set-2: $N_\mathrm{f} = 55$ & $\beta= (0,0.40]$ & $\sigma_0=[0.0134,0.0170]$ & $\left(\alpha_+/\alpha_-\right)^{\beta} \geq 7.5$\\
   \hline
     (3) Set-3: $N_\mathrm{f} = 60$ & $\beta= (0,0.40]$ & $\sigma_0=[0.0135,0.0170]$ & $\left(\alpha_+/\alpha_-\right)^{\beta} \geq 7.5$\\
   \hline
     \hline
  \end{tabular}%
 }
  \caption{Viable ranges on entropic parameters coming from the inflationary phenomenology for three different choices of $N_\mathrm{f}$.}
  \label{Table-1}
 \end{table}

It may be noticed that the viable ranges of the entropic parameters do not change drastically with the total e-fold of the inflationary era.

Thus as a whole, the entropic cosmology corresponding to the $S_\mathrm{g}$ (with varying $\gamma = \gamma(N)$) triggers a viable inflation with an exit, provided the entropic parameters satisfy the constraints shown in the above table. At this stage it is important to explore the post inflationary phase and its route to the Standard Big-Bang Cosmology in the present context of generalized entropic scenario.

\subsection{From inflation to reheating}\label{sec-reheating}

After the end of inflation, the energy density originated from the $S_\mathrm{g}$ decays to relativistic particles with a certain decay width ($\Gamma$) that is generally considered to be a constant. In particular, we will consider perturbative reheating caused due to a coupling between entropic energy and relativistic particles, in the same spirit of \cite{Dai:2014jja,Cook:2015vqa}. By using Eq.(\ref{solution-viable-inf-2}) and Eq.(\ref{FRW-eq-viable-inf-main1}), we obtain the entropic energy density and the corresponding EoS as,
\begin{eqnarray}
 \rho_\mathrm{g} = \frac{3}{16G^2} \left(\frac{\alpha_+}{\beta}\right)
\left[\frac{2\exp{\left[-\int^{N}\sigma(N)dN\right]}}
{1 + \sqrt{1 + 4\left(\alpha_+/\alpha_-\right)^{\beta}\exp{\left[-2\int^{N}\sigma(N)dN\right]}}}\right]^{1/\beta}~~,
 \label{reh-1}
\end{eqnarray}
and
\begin{eqnarray}
 w_\mathrm{g} = \frac{p_\mathrm{g}}{\rho_\mathrm{g}} = -1 + \left(\frac{16G^2}{9}\right)\rho_\mathrm{g}~\sigma(N)
 \left[\frac{\left(1 + \frac{\pi\alpha_+}{\beta GH^2}\right)^{\beta} - \left(1 + \frac{\pi\alpha_-}{\beta GH^2}\right)^{-\beta}}
{\alpha_+\left(1 + \frac{\pi\alpha_+}{\beta GH^2}\right)^{\beta-1} + \alpha_-\left(1 + \frac{\pi\alpha_-}{\beta GH^2}\right)^{-\beta-1}}\right]~~,
 \label{reh-3}
\end{eqnarray}
respectively. Since $\rho_\mathrm{g}$ decays to radiation energy density during the reheating stage, the effective EoS during the same is given by,
\begin{eqnarray}
 w_\mathrm{eff} = \frac{3w_\mathrm{g}\rho_\mathrm{g} + \rho_\mathrm{R}}{3\left(\rho_\mathrm{g} + \rho_\mathrm{R}\right)}~~,
 \label{R2-1}
\end{eqnarray}
where $\rho_\mathrm{R}$ represents the radiation energy density. During the perturbative reheating, the Hubble parameter is generally much larger than the decay width (i.e. $H \gg \Gamma$), due to which, the comoving entropic energy density remains conserved with the cosmic expansion of the universe. However with time, the Hubble parameter eventually becomes comparable to the decay width (i.e. $H \sim \Gamma$), when the entropic energy density instantaneously decays to the radiation. This indicates the end of reheating, in particular, the reheating ends when $H = \Gamma$ satisfies. Based on these arguments, the effective EoS during the reheating era can be expressed by,
\begin{align}
w_\mathrm{eff}=
\begin{cases}
w_\mathrm{g}~~;~~~\mathrm{during~the ~reheating}~, & \\
1/3~~;~~~\mathrm{at~the~end~of~reheating}~,
\end{cases}
\label{R2-2}
\end{align}
and the end of reheating gets continuously connected to the radiation dominated era. Moreover, in analogy of standard scalar field cosmology, we consider that the Hubble parameter during the reheating stage evolves as a power law, in particular,
\begin{eqnarray}
 H(N) = H_\mathrm{f}~\mathrm{exp}\left[-\left(N-N_\mathrm{f}\right)/m\right]~~,
 \label{nreh1}
\end{eqnarray}
during the reheating era (in terms of e-folding number), where $m$ is the exponent. Here we need to mention that $N = N_\mathrm{f}$ at the end of inflation and $N = N_\mathrm{f} + N_\mathrm{re}$ when the reheating ends, where $N_\mathrm{f}$ and $N_\mathrm{re}$ represent the duration of inflation and reheating respectively. The above evolution of $H(N)$ immediately leads to the EoS during the reheating as,
\begin{eqnarray}
 w_\mathrm{eff} = -1 - \frac{2}{3}\frac{d\mathrm{ln}H}{dN} = -1 + \frac{2}{3m}
 \label{reh-5}
\end{eqnarray}
which is actually a constant, resulting to a 'perfect fluid' nature of the entropic energy during the reheating stage. In order to confront the ansatz of the Hubble parameter in Eq.~(\ref{nreh1}) with the governing Friedmann Eq.(\ref{FRW-eq-viable-inf-main1}), we need to find the form of $\sigma(N)$ during the reheating stage in such a way that the Hubble parameter follows Eq.(\ref{nreh1}); for this purpose, we use the ansatz of $H(N)$ from Eq.~(\ref{nreh1}) to Eq.(\ref{FRW-eq-viable-inf-main1}) and the reconstructed $\sigma(N)$ is obtained as,
\begin{eqnarray}
 \sigma(N) = \left(\frac{2\pi}{G}\right)\frac{e^{2\left(N - N_\mathrm{f}\right)/m}}{mH_\mathrm{f}^2}
 \left[\frac{\alpha_+\left(1 + \frac{\pi\alpha_+}{\beta GH_\mathrm{f}^2}~e^{2\left(N - N_\mathrm{f}\right)/m}\right)^{\beta-1} + \alpha_-\left(1 + \frac{\pi\alpha_-}{\beta GH_\mathrm{f}^2}~e^{2\left(N - N_\mathrm{f}\right)/m}\right)^{-\beta-1}}
{\left(1 + \frac{\pi\alpha_+}{\beta GH_\mathrm{f}^2}~e^{2\left(N - N_\mathrm{f}\right)/m}\right)^{\beta} - \left(1 + \frac{\pi\alpha_-}{\beta GH_\mathrm{f}^2}~e^{2\left(N - N_\mathrm{f}\right)/m}\right)^{-\beta}}\right]\nonumber\\
 ~~~~~~;~~\mathrm{during~reheating}~~.
 \label{nreh2}
\end{eqnarray}
Thus as a whole, $\gamma(N)$ is of the form of Eq.(\ref{gamma function}) where $\sigma(N)$ during the inflation and during the reheating stages are given by Eq.(\ref{sigma function}) and Eq.(\ref{nreh2}) respectively. Clearly the functional behaviour of $\gamma(N)$ during inflation is different than that of during the reheating stage, and moreover, the change is continuous at the junction of inflation-to-reheating. This in turn reveals the continuous change of the Hubble parameter from a quasi dS evolution during inflation to a power law evolution during the reheating era.

\begin{figure}[!h]
\begin{center}
\centering
\includegraphics[width=3.0in,height=2.0in]{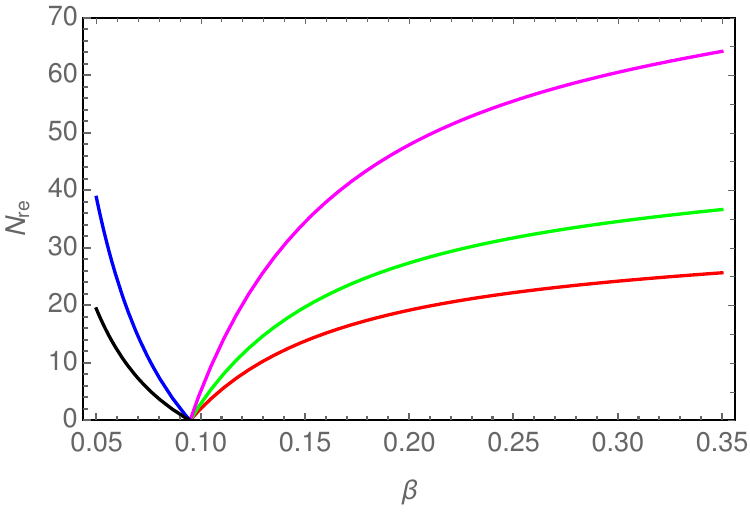}
\includegraphics[width=3.0in,height=2.0in]{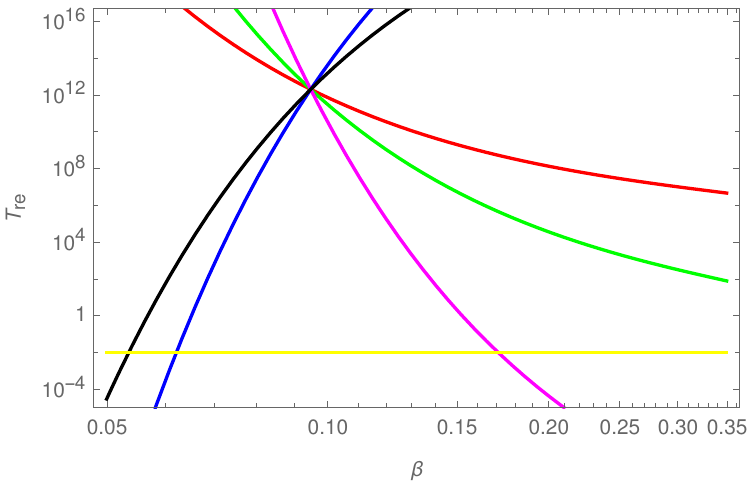}
\caption{{\color{blue}Left Plot}: $N_\mathrm{re}$ vs. $\beta$; {\color{blue}Right Plot}: $T_\mathrm{re}$ vs. $\beta$ for $N_\mathrm{f} = 50$ and various values of $w_\mathrm{eff}$. In both the plots, we consider $\sigma_0 = 0.015$. The reheating EoS parameter is taken as $w_\mathrm{eff} = 0 ~(\mathrm{Red~curve}), 0.1~(\mathrm{Green~curve}), 0.2~(\mathrm{Magenta~curve}), \frac{2}{3}~(\mathrm{Blue~curve}), 1~(\mathrm{Black~curve})$ respectively. Moreover the yellow curve in the right plot is the BBN temperature $\sim 10^{-2}\mathrm{GeV}$.}
\label{plot-Reh1}
\end{center}
\end{figure}

The reheating era is generally parametrized by its total e-fold number ($N_\mathrm{re}$) and the reheating temperature ($T_\mathrm{re}$), which are constrained by $N_\mathrm{re} > 0$ and $T_\mathrm{re} > T_\mathrm{BBN} \sim 10^{-2}\mathrm{GeV}$ that, in turn, will put viable constraints on the entropic parameters. Following \cite{Odintsov:2023vpj}, we write $N_\mathrm{re}$ and $T_\mathrm{re}$ (in terms of the entropic parameters) in the present context of generalized entropic cosmology as follows:
\begin{align}
 N_\mathrm{re}&=\frac{4}{\left(1 - 3w_\mathrm{eff}\right)}\times\nonumber\\
 &\Bigg\{61.6 - \frac{1}{4\beta}\ln{\left[\frac{\beta\mathrm{e}^{-\left(1 + \sigma_0N_\mathrm{f}\right)}\left\{1 + \sqrt{1 + \mathrm{e}^{2\left(1 + \sigma_0N_\mathrm{f}\right)}\left[\left(\frac{1+\sigma_0}{2\beta}\right)^2 - 1\right]}\right\}^2}
{\left(16\pi^2\alpha_+/3\beta\right)^{\beta}~\left\{1+\sigma_0 + 2\beta\right\}}\right]} - N_\mathrm{f}\Bigg\}
 \label{et-10}
\end{align}
and
\begin{eqnarray}
 T_\mathrm{re} = H_\mathrm{i}\left(\frac{43}{11g_\mathrm{re}}\right)^{1/3}\left(\frac{T_0}{k/a_0}\right)e^{-\left(N_\mathrm{f} + N_\mathrm{re}\right)}~~,
 \label{et-4}
\end{eqnarray}
respectively. Here $\frac{k}{a_0} = 0.05\mathrm{Mpc}^{-1}$ is the CMB scale and $T_0 = 2.93\mathrm{K}$ symbolizes the present temperature of the universe. Moreover the quantity $H_\mathrm{i}$, i.e the Hubble parameter at the beginning of inflation, is obtained from Eq.(\ref{solution-viable-inf-2}) (with $\sigma(N)$ is given by Eq.(\ref{sigma function})) as,
\begin{eqnarray}
 H_\mathrm{i} = 4\pi M_\mathrm{Pl}\sqrt{\frac{\alpha_+}{\beta}}
\left[\frac{2}
{\left\{1 + \sqrt{1 + \mathrm{exp}\left[2\left(1 + \sigma_0N_\mathrm{f}\right)\right]\left[\left(\frac{1+\sigma_0}{2\beta}\right)^2 - 1\right]}\right\}}\right]^{1/(2\beta)}~~.
 \label{et-8}
\end{eqnarray}
Owing to the constraints of $N_\mathrm{re} > 0$ and $T_\mathrm{re} > T_\mathrm{BBN} \approx 10^{-2}\mathrm{GeV}$, the above theoretical expectations of $N_\mathrm{re}$ and $T_\mathrm{re}$ in the present context in turn put certain constraints on the entropic parameters. These, along with inflationary phenomenology in account, are presented in the Table[\ref{Table-0}]. For instance, here we give the plot of $N_\mathrm{re}$ and $T_\mathrm{re}$ w.r.t. the parameter $\beta$ for $N_\mathrm{f} = 50$, see Fig.~[\ref{plot-Reh1}].

\begin{table}[h]
  \centering
 {%
  \begin{tabular}{|c|c|c|c|}
   \hline
    Viable choices of $N_\mathrm{f}$ & Viable range of $\beta$ & Viable range of $\left(\frac{\alpha_+}{\alpha_-}\right)^{\beta}$ & Reheating EoS parameter\\

   \hline
  \hline
   (1) Set-1: $N_\mathrm{f} = 50$ & (a) $0.05 < \beta < 0.10$ & $2\times10^{5} < \left(\frac{\alpha_+}{\alpha_-}\right)^{\beta} < 8.5\times10^{5}$ & $\frac{1}{3} < w_\mathrm{eff} < 1$\\
   \hline
     & (b) $0.10 < \beta < 0.35$ & $7.5 < \left(\frac{\alpha_+}{\alpha_-}\right)^{\beta} < 2\times10^{5}$ & $-\frac{1}{3} < w_\mathrm{eff} < \frac{1}{3}$\\
     \hline
    (2) Set-2: $N_\mathrm{f} = 55$ & (a) $0.06 < \beta < 0.22$ & $4\times10^{4} < \left(\frac{\alpha_+}{\alpha_-}\right)^{\beta} < 5\times10^{5}$ & $\frac{1}{3} < w_\mathrm{eff} < 1$\\
   \hline
     & (b) $0.22 < \beta < 0.40$ & $7.5 < \left(\frac{\alpha_+}{\alpha_-}\right)^{\beta} < 4\times10^{4}$ & $-\frac{1}{3} < w_\mathrm{eff} < \frac{1}{3}$\\
     \hline
     (3) Set-3: $N_\mathrm{f} = 60$ & (a) $0.08 < \beta < 0.40$ & $7.5 < \left(\frac{\alpha_+}{\alpha_-}\right)^{\beta} < 3\times10^{5}$ & $\frac{1}{3} < w_\mathrm{eff} < 1$\\
   \hline
     \hline
  \end{tabular}%
 }
  \caption{Viable ranges on entropic parameters coming from both the inflation and reheating phenomenology for three different choices of $N_\mathrm{f}$. In each of the set, the parameter $\sigma_0$ lies within the range $0.013 \lesssim \sigma_0 \lesssim 0.0170$ which arises from the inflationary constraints and does not change from reheating phenomenology.}
  \label{Table-0}
 \end{table}

In addition, we would like to mention that the reheating EoS parameter lies in the range $w_\mathrm{eff} > \frac{1}{3}$ for the cases $N_\mathrm{f} \gtrsim 57.3$, which is also reflected from the above Table.~[\ref{Table-0}] in Set-3. Thus we may notice that the entropic parameters corresponding to $S_\mathrm{g}$ gets further constrained by the input of the reheating stage.

As a whole --- the horizon cosmology, based on the 4-parameter generalized entropy of the apparent horizon, proves to be useful in explaining the inflation-to-reheating of the early universe and its route to the radiation dominated era. Owing to the fact that the entropic energy density decays to the relativistic particles (behaves as normal matter fields inside the horizon) at the end of reheating, the entropy of the apparent horizon acquires the form of the Bekenstein-Hawking entropy from the end of the reheating. Actually the 4-parameter generalized entropy has a certain evolution with cosmic time, and consequently, it transforms to the Bekenstein-Hawking entropy at the end of reheating. This ensures the usual evolution of the universe during the radiation dominated era. Therefore we may argue that the standard cosmic evolution during the radiation era is a consequence of the 4-parameter generalized entropy and its evolution from the early stage of the universe.

\section{Evolution and spectrum of primordial GWs generated during inflation}
In this section we will address the evolution, and consequently, the spectrum of primordial GWs generated during inflation. For the purpose of the spectrum, we will primarily focus on the dimensionless energy density of the GWs.

\subsection{Generation and evolution of GWs during inflation}
Let $h_{ij}(t,\vec{x})$ be the tensor perturbation characterizing GWs over a spatially flat FLRW spacetime, and hence, the spacetime metric can be expressed by,
\begin{eqnarray}
 ds^2 = -dt^2 + a^2(t)\left[\left(\delta_{ij} + h_{ij}\right)dx^{i}dx^{j}\right]~~.
 \label{I-1}
\end{eqnarray}
In absence of any anisotropic source (which is indeed the case in the present context), the tensor perturbation variable is governed by the following equation:
\begin{eqnarray}
 \ddot{h}_{ij} + 3H\dot{h}_{ij} - \nabla^2h_{ij} = 0~~,
 \label{I-2}
\end{eqnarray}
where $\nabla^2$ represents the Laplacian operator. On quantizing the tensor perturbation, $h_{ij}(t,\vec{x})$ can be decomposed in terms of its Fourier modes $h(k,t)$ as:
\begin{eqnarray}
 \hat{h}_{ij}(t,\vec{x}) = \sum_{\lambda=+,\times}\int \frac{d^3\vec{k}}{(2\pi)^{3/2}}\left[\hat{a}_k^{\lambda}\epsilon_{ij}^{\lambda}(\vec{k})h(k,t)\mathrm{e}^{i\vec{k}.\vec{x}} + c.c.\right]~~,
 \label{I-3}
\end{eqnarray}
where $\lambda = +,\times$ denotes two types of polarizations of the GWs, $\epsilon_{ij}^{\lambda}(\vec{k})$ are the polarization tensor and $\hat{a}_k$ ($\hat{a}_k^{+}$) are the annihilation (creation) operators respectively that satisfy the usual commutation rules. Moreover, from the transverse condition of GWs, i.e. due to $\partial_{i}h^{ij} = 0$, one immediately gets $k^{i}\epsilon_{ij}^{\lambda}(\vec{k}) = 0$. Owing to Eq.~(\ref{I-2}), $h(k,t)$ obeys,
\begin{eqnarray}
 \ddot{h}(k,t) + 3H\dot{h}(k,t) + \frac{k^2}{a^2}h(k,t) = 0~~.
 \label{I-4}
\end{eqnarray}
Consequently, the tensor power spectrum is defined by two point correlations of $h(k,t)$ over a suitable vacuum state generally considered to be the Bunch-Davies vacuum state. Passing to the conformal time $\eta$ (defined by $d\eta = dt/a(t)$) and introducing the Mukhanov-Sasaki variable as
\begin{equation}
    v(k,\eta) = \frac{a}{2} M_\mathrm{Pl}h(k,\eta) \ ,
    \label{I-5}
\end{equation}
Eq.~(\ref{I-4}) transforms to,
\begin{equation}\label{EoM_MS_variable}
    v''(k,\eta)+\left(k^2-\frac{a''}{a}\right)v(k,\eta)=0 \ ,
\end{equation}
where the overprime denotes $\frac{d}{d\eta}$. Clearly the Mukhanov-Sasaki variable obeys an oscillator like equation with a time dependent mass term, due to which, the particle production corresponding to the quantized tensor perturbation (over the Bunch-Davies vacuum) will occur, and that may be accounted through the GWs spectrum.

As demonstrated in Sec.~[\ref{sec-inf}], the entropic cosmology with the 4-parameter generalized entropy for the apparent horizon is able to trigger a viable quasi-dS inflation that is well consistent with the Planck data, and moreover, the inflation has an exit at around $N_\mathrm{f} = 60$ e-fold number. Since the slow roll parameter remains almost constant (and nearly equal to zero) during most of the inflationary period (the change in the slow roll parameter occurs near the end of inflation), we can safely consider the scale factor during the inflation to be of the de-Sitter form, i.e. $a(\eta) = -\frac{1}{H_\mathrm{i}\eta}$, where $H_\mathrm{i}$ is the Hubble parameter during inflation. With this form of $a(\eta)$, Eq.(\ref{EoM_MS_variable}) becomes
\begin{equation}\label{EoM_MS_variable_Inflation}
    v''(k,\eta)+\left(k^2-\frac{2}{\eta^2}\right)v(k,\eta)=0 \ ,
\end{equation}
on solving which, we get,
\begin{equation}\label{MS_variable_Inflation}
    v(k,\eta) = \frac{e^{-ik\eta}}{\sqrt{2k}}\left(1-\frac{i}{k\eta}\right) \ .
\end{equation}
This solution is compatible with the Bunch-Davies initial condition in the deep inside of the Hubble radius, given by,
\begin{equation}
    \lim_{|k\eta|\gg 1}v(k,\eta) = \frac{1}{\sqrt{2k}}e^{-ik\eta} \ .
    \label{I-6}
\end{equation}
%\begin{figure}
 %   \centering
  %  \includegraphics[scale=.5]{evolution.pdf}
   % \caption{Caption}
    %\label{fig:enter-label}
%\end{figure}
Plugging the above solution of $v(k,\eta)$ into Eq.~(\ref{I-5}) yields the solution of $h(k,\eta)$ as
\begin{align}
    h(k,\eta) = -\sqrt{\frac{2}{k}}\left(\frac{H_\mathrm{i}}{M_\mathrm{Pl}}\right)\eta \,  e^{-ik\eta}\left(1-\frac{i}{k\eta}\right) \label{h_inflation} \ .
\end{align}
Consequently we calculate
\begin{equation}\label{I-7}
    h'(k,\eta)=i \, \sqrt{\frac{2}{k}} \left(\frac{H_\mathrm{i}}{M_\mathrm{Pl}}\right)(k\eta)e^{-ik\eta}
\end{equation}
The expressions for $h(k,\eta)$ and $h'(k,\eta)$ at the end of inflation will act as initial conditions for the reheating era. If $\eta=\eta_\mathrm{f}$ represents the end of inflation, then
\begin{equation}\label{k_eta_f}
    -k\eta_\mathrm{f}=\left(\frac{k}{k_\mathrm{CMB}}\right)e^{-N_\mathrm{f}} \ ,
\end{equation}
where $k_\mathrm{CMB} \sim 0.05\mathrm{Mpc}^{-1}$ is the large scale mode whose horizon crossing instant is considered to be the beginning of inflation (i.e. at $N= 0$), and recall that $N_\mathrm{f}$ is the total e-fold for the inflationary period. Moreover to arrive at the above expression, we use $\eta_\mathrm{f} = -1/\left(a_\mathrm{f}H_\mathrm{i}\right)$ with $a_\mathrm{f}$ being the scale factor at the end of inflation. For $k<k_\mathrm{f}$, where $k_\mathrm{f}$ is the mode that crosses the horizon at the end of inflation, Eq.(\ref{k_eta_f}) can be approximated by $|k\eta_\mathrm{f}| \rightarrow 0$. Here it deserves mentioning that we will restrict ourselves to the modes satisfying $k < k_\mathrm{f}$ to determine the GWs spectrum in the present work. Due to $|k\eta_\mathrm{f}| \rightarrow 0$, we have the following expressions for $h(k,\eta)$ and $h'(k,\eta)$ at the end of inflation:
\begin{eqnarray}
 h(k,\eta_\mathrm{f}) \approx i\frac{\sqrt{2}}{k^{3/2}}\left(\frac{H_\mathrm{i}}{M_\mathrm{Pl}}\right)~~~~~~~~\mathrm{and}~~~~~~~~~~h'(k,\eta_\mathrm{f}) \approx 0~~,
 \label{I-8}
\end{eqnarray}
respectively, which will be useful for the continuity conditions of the tensor perturbation at the junction of inflation-to-reheating. Moreover $h(k,\eta_\mathrm{f})$ may be equivalently argued as the tensor perturbation amplitude at super-Hubble scale (i.e. for $|k\eta| \ll 1$ where $h(k,\eta)$ gets frozen).

Before moving further, let us introduce the transfer function $\chi(k,\eta)$ as
\begin{equation}\label{I-9}
    h(k,\eta)=\left[\lim_{|k\eta|\ll 1}h(k,\eta)\right]\chi(k,\eta) = i\sqrt{\frac{2}{k^3}}\left(\frac{H_\mathrm{i}}{M_\mathrm{Pl}}\right)\chi(k,\eta) \ ,
\end{equation}
which is helpful to follow the post-inflationary evolution of the tensor perturbation. It may be noted that the transfer function is connected with $h(k,\eta)$ by the amplitude of the tensor perturbation at super-Hubble scale.

\subsection{Evolution of GWs during reheating}
After the inflation ends, the universe enters to the reheating phase when the entropic energy density decays to relativistic particles, as demonstrated in Sec.~[\ref{sec-reheating}]. The transfer function during the reheating era obeys,
\begin{equation}\label{EoM_chi_general}
    \Ddot{\chi}(k,t)+3H\dot{\chi}(k,t)+\frac{k^2}{a^2}\chi(k,t)=0 \ .
\end{equation}
where $H$ represents the Hubble parameter during the same. It is more convenient to change the variable of the above differential equation from cosmic time ($t$) to scale factor ($a$). As a result, Eq.~(\ref{EoM_chi_general}) turns out to be
\begin{equation}\label{Reh-1}
    a^4H^2\frac{d^2\chi}{da^2}+a^4H\frac{dH}{da}\frac{d\chi}{da}+4a^3H^2\frac{d\chi}{da}+k^2\chi=0 \ .
\end{equation}
Introducing $A=\frac{a}{a_\mathrm{f}}$ and with this new redefined scale factor, the above equation of motion for $\chi(k,A)$ becomes
\begin{equation}\label{EoM_chi_A}
    \frac{d^2\chi}{dA^2}+\left(\frac{1}{H}\frac{dH}{dA}+\frac{4}{A}\right)\frac{d\chi}{dA}+\frac{k^2}{a_\mathrm{f}^2A^4H^2}\chi=0 \ ,
\end{equation}
note that at the end of inflation we have $A=1$. The Hubble parameter during the reheating stage can be immediately written from Eq.~(\ref{nreh1}) as,
\begin{equation}
    H \propto A^{-\frac{3}{2}(1+w_\mathrm{eff})} \ ,
\end{equation}
where recall that $w_\mathrm{eff}$ is the reheating EoS parameter originated from the entropic energy density of the 4-parameter generalized entropy.
By substituting the above expression of $H=H(A)$ into Eq.~(\ref{EoM_chi_A}) and using $k_\mathrm{f}=a_\mathrm{f}H_\mathrm{f}$, we have, during reheating,
\begin{equation}\label{Reh-2}
    \frac{d^2\chi}{dA^2}+\frac{1}{2A}(5-3w_\mathrm{eff})\frac{d\chi}{dA}+\frac{(k/k_\mathrm{f})^2}{A^{1-3w_\mathrm{eff}}}\chi = 0 \ .
\end{equation}
The continuity conditions of tensor perturbation at the junction of inflation-to-reheating are shown in Eq.~(\ref{I-8}) which, in terms of the transfer function, leads to the following conditions at $A=1$:
\begin{equation}\label{Reh-3}
    \chi(k,A=1) = 1 \hspace{0.5cm} \text{and} \hspace{0.5cm}\frac{d\chi}{dA}\bigg|_{A=1}=0
\end{equation}
respectively. With these two continuity conditions, we solve Eq.~(\ref{Reh-2}) for $\chi(k,A)$ during the reheating phase, and is obtained as,
\begin{equation}\label{chi_reheating}
    \chi^\mathrm{RH}(k,A)=\frac{\mathscr{N}(k,A)}{\mathscr{D}(k)} \ .
\end{equation}
Here, along with
\begin{eqnarray}
 \nu \equiv \frac{3w_\mathrm{eff}-3}{6w_\mathrm{eff}+2} = \frac{1-3m}{2+2m}~~,
 \label{Reh-4}
\end{eqnarray}
( where in the second equality we use Eq.~(\ref{reh-5}), one may recall that $m$ appears from the varying entropic parameter $\gamma(N)$ that actually ensures the continuous transition of the Hubble parameter from a quasi-dS inflationary era to a power law evolution during the reheating, see the discussion after Eq.~(\ref{nreh2}) ), the quantities $\mathscr{N}(k,A)$ and $\mathscr{D}(k)$ are given by
\begin{align}\label{chi_reheating_D}
    \mathscr{D}(k)\equiv2k\Bigg\{&J_{\nu} \left(\frac{2 k/k_\mathrm{f}}{1+3w_\mathrm{eff}}\right)\left[J_{-\nu-1} \left(\frac{2 k/k_\mathrm{f}}{1+3w_\mathrm{eff}}\right)-J_{-\nu+1} \left(\frac{2 k/k_\mathrm{f}}{1+3w_\mathrm{eff}}\right)\right]\\
    -&J_{-\nu} \left(\frac{2 k/k_\mathrm{f}}{1+3w_\mathrm{eff}}\right)\left[J_{\nu-1} \left(\frac{2 k/k_\mathrm{f}}{1+3w_\mathrm{eff}}\right)-J_{\nu+1} \left(\frac{2 k/k_\mathrm{f}}{1+3w_\mathrm{eff}}\right)\right]
    \Bigg\}
\end{align}
and
\begin{equation}\label{chi_reheating_N}
    \mathscr{N}(k,A)\equiv A^{-\frac{3+3w_\mathrm{eff}}{4}}\left[J_\nu\left(\frac{2k/k_\mathrm{re}}{1+3w_\mathrm{eff}}\left(\frac{A}{A_\mathrm{re}}\right)^\frac{1+3w_\mathrm{eff}}{2}\right) \mathscr{N}_1(k)+J_{-\nu}\left(\frac{2k/k_\mathrm{re}}{1+3w_\mathrm{eff}}\left(\frac{A}{A_\mathrm{re}}\right)^\frac{1+3w_\mathrm{eff}}{2}\right) \mathscr{N}_2(k) \right] \ ,
\end{equation}
respectively, with
\begin{equation}\label{Reh-5}
    \mathscr{N}_1(k)\equiv 2kJ_{-\nu-1}\left(\frac{2k/k_\mathrm{f}}{1+3w_\mathrm{eff}}\right)-2kJ_{-\nu+1}\left(\frac{2k/k_\mathrm{f}}{1+3w_\mathrm{eff}}\right)+3k_\mathrm{f}(w_\mathrm{eff}-1)J_{-\nu}\left(\frac{2k/k_\mathrm{f}}{1+3w_\mathrm{eff}}\right)
\end{equation}
and
\begin{equation}\label{Reh-6}
    \mathscr{N}_2(k)\equiv 2kJ_{\nu+1}\left(\frac{2k/k_\mathrm{f}}{1+3w_\mathrm{eff}}\right)-2kJ_{\nu-1}\left(\frac{2k/k_\mathrm{f}}{1+3w_\mathrm{eff}}\right) - 3k_\mathrm{f}(w_\mathrm{eff}-1)J_{\nu}\left(\frac{2k/k_\mathrm{f}}{1+3w_\mathrm{eff}}\right) \ .
\end{equation}
We will eventually use this solution of $\chi(k,A)$ to arrive at the GWs spectrum at present epoch. At the moment, we also determine $d\chi^\mathrm{RH}(k,A)/dA$ from Eq.~(\ref{chi_reheating}), which is given by,
\begin{eqnarray}
 \frac{d\chi^\mathrm{RH}(k,A)}{dA} = \frac{1}{\mathscr{D}(k)}\frac{d\mathscr{N}(k,A)}{dA}~~,
 \label{Reh-7}
\end{eqnarray}
(where one may use the property for the derivative of the Bessel function: $J'_{\nu}(x) = \frac{1}{2}\left[J_{\nu-1}(x) - J_{\nu+1}(x)\right]$). At the end of this subsection, we would like to mention that the above expressions of $\chi^\mathrm{RH}(k,A)$ and $d\chi^\mathrm{RH}(k,A)/dA$ at $A = A_\mathrm{re}$ provide the initial conditions of the transfer function at radiation dominated era ($A_\mathrm{re}$ symbolizes the instant of the end of the reheating).

\subsection{GWs' evolution during radiation era and its spectrum today}

During the radiation domination (RD), the Hubble parameter follows the evolution $H \propto a^{-2}$ with the scale factor, and thus the Hubble parameter during the RD may be connected with that of at the end of the reheating as,
\begin{eqnarray}
 A^4H^2 = A_\mathrm{re}^4H_\mathrm{re}^2~~.
 \label{rd-1}
\end{eqnarray}
Using this, the equation of motion for $\chi(k,A)$ during RD takes the following form
\begin{equation}\label{EoM_chi_RD}
    \frac{d^2\chi}{dA^2}+\frac{2}{A}\frac{d\chi}{dA}+\frac{(k/k_\mathrm{re})^2}{A_{re}^2}\chi(k,A)=0~~,
\end{equation}
where we may recall that $k_\mathrm{re}=a_\mathrm{re}H_\mathrm{re}$ is the mode that re-enters the horizon at the end of the reheating. On solving Eq.~(\ref{EoM_chi_RD}), we get the transfer function during RD as follows:
\begin{equation}\label{chi_RD}
    \chi^\mathrm{RD}(k,A) = \frac{1}{A}\left(c_\mathrm{1}\mathrm{e}^{-ibA} + c_\mathrm{2}\mathrm{e}^{ibA}\right) \ ,
\end{equation}
with $b = \left(\frac{k}{k_\mathrm{re}}\right)\frac{1}{A_\mathrm{re}}$. Moreover $c_\mathrm{1}$ and $c_\mathrm{2}$ are the integration constants that can be determined from the continuity conditions of the transfer function at the junction of reheating-to-radiation, given by:
\begin{eqnarray}
 \chi^\mathrm{RD}(k,A_\mathrm{re}) = \chi^\mathrm{RH}(k,A_\mathrm{re})~~~~~~~~~~~\mathrm{and}~~~~~~~~~~
 \frac{d\chi^\mathrm{RD}(k,A)}{dA}\bigg|_{A_\mathrm{re}} = \frac{d\chi^\mathrm{RH}(k,A)}{dA}\bigg|_{A_\mathrm{re}}~~,
 \label{rd-2}
\end{eqnarray}
where the quantities at the end of the reheating (giving by the suffix 're') can be determined from Eq.~(\ref{chi_reheating}). Such continuity conditions lead to the integration constants as,
%\begin{subequations}\label{c_i}
\begin{align}\label{c_i}
    c_\mathrm{1,2}=\frac{e^{ibA_\mathrm{re}}}{2}\left[\left(A_\mathrm{re} \mp \frac{1}{ib}\right)\chi^\mathrm{RH}(k,A_\mathrm{re}) \mp \left(\frac{A_\mathrm{re}}{ib}\right)\frac{d\chi^\mathrm{RH}(k,A_\mathrm{re})}{dA}\right] \ .
\end{align}
%\end{subequations}
Consequently the final form of the transfer function during RD is given by,
\begin{eqnarray}\label{rd-3}
    \chi^\mathrm{RD}(k,A)&=&\frac{e^{-ib(A-A_\mathrm{re})}}{2A}\left[\left(A_\mathrm{re} - \frac{1}{ib}\right)\chi^\mathrm{RH}(k,A_\mathrm{re}) - \left(\frac{A_\mathrm{re}}{ib}\right)\frac{d\chi^\mathrm{RH}(k,A_\mathrm{re})}{dA}\right]\nonumber\\
    &+&\frac{e^{ib(A-A_\mathrm{re})}}{2A}\left[\left(A_\mathrm{re} + \frac{1}{ib}\right)\chi^\mathrm{RH}(k,A_\mathrm{re}) + \left(\frac{A_\mathrm{re}}{ib}\right)\frac{d\chi^\mathrm{RH}(k,A_\mathrm{re})}{dA}\right]  \ .
\end{eqnarray}

Having obtained $\chi^\mathrm{RD}(k,A)$, let us now focus to the observable quantity of our interest, in particular, to the dimensionless energy density of GWs. As mentioned earlier that we are interested over the modes which re-enter the horizon during the epochs of reheating and radiation domination, i.e., over $k<k_\mathrm{f}$. The dimensionless energy density of GWs, at an instant $A$, is defined by
\begin{align}\label{rd-4}
    \Omega_\mathrm{GW}(k,A)=&\frac{1}{3H^2M_\mathrm{Pl}^2}\rho_\mathrm{GW}(k,A) \\
    =&\frac{1}{6A^2a_\mathrm{f}^2H^2}\left(\frac{k^3}{2\pi^2}\right)\left\{a_\mathrm{f}^2A^4H^2\left|\frac{dh(k,A)}{dA}\right|^2 + k^2|h(k,A)|^2\right\} \ ,
\end{align}
where $\rho_\mathrm{GW}(k,A)$ represents the GWs energy density (at the instant $A$) per logarithmic interval of the modes. In terms of the transfer function, $\Omega_\mathrm{GW}(k,A)$ from Eq.~(\ref{rd-4}) becomes,
\begin{equation}\label{rd-5}
    \Omega_\mathrm{GW}(k,A)=\frac{1}{6\pi^2}\left(\frac{H_\mathrm{i}}{M_\mathrm{Pl}}\right)^2\left\{A^2\left|\frac{d\chi(k,A)}{dA} \right|^2 + \frac{k^2}{a^2H^2}|\chi(k,A)|^2\right\} \ ,
\end{equation}
with $H_\mathrm{i}$ being the inflationary Hubble parameter, see Eq.~(\ref{et-8}). We will eventually determine $\Omega_\mathrm{GW}(k,A)$ around the present epoch when the Hubble parameter follows the evolution like $H\propto A^{-2}$ (i.e. of the RD era), and the transfer function is given by Eq.~(\ref{chi_RD}). As a result, one may simplify the term $k/(aH)$ (present in the r.h.s. of Eq.~(\ref{rd-5})) as,
\begin{eqnarray}
 \frac{k}{aH} = b\left(\frac{k_\mathrm{re}A_\mathrm{re}}{aH}\right) = b\left(\frac{A_\mathrm{re}^2H_\mathrm{re}}{AH}\right) = b\left(\frac{A^2H}{AH}\right) = bA~~,
 \label{rd-6}
\end{eqnarray}
where $b = \left(\frac{k}{k_\mathrm{re}}\right)\frac{1}{A_\mathrm{re}}$ (see after Eq.~(\ref{chi_RD})), and we have used $H\propto A^{-2}$ in the third equality. Therefore $\Omega_\mathrm{GW}(k,A)$ from Eq.~(\ref{rd-5}) can be expressed by,
\begin{equation}\label{rd-7}
    \Omega_\mathrm{GW}(k,A)=\frac{1}{6\pi^2}\left(\frac{H_\mathrm{i}}{M_\mathrm{Pl}}\right)^2\left\{A^2\left|\frac{d\chi^\mathrm{RD}(k,A)}{dA} \right|^2 + b^2A^2\left|\chi^\mathrm{RD}(k,A)\right|^2\right\} \ .
\end{equation}
The modes of interest are well inside the Hubble radius at late times (say, close to the radiation-matter equality) during radiation domination. Consequently the dimensionless energy density parameter $\Omega_\mathrm{GW}^\mathrm{(0)}(k)$ today (i.e. at present epoch) is given by,
\begin{eqnarray}
 \Omega_\mathrm{GW}^\mathrm{(0)}(k)h^2&\simeq&\left(\frac{g_{r,0}}{g_{r,eq}}\right)^{1/3}\Omega_\mathrm{R}h^2\Omega_\mathrm{GW}(k,A)\nonumber\\
 &=&\frac{1}{6\pi^2}\left(\frac{g_{r,0}}{g_{r,eq}}\right)^{1/3}\Omega_\mathrm{R}h^2\left(\frac{H_\mathrm{i}}{M_\mathrm{Pl}}\right)^2\left\{A^2\left|\frac{d\chi^\mathrm{RD}(k,A)}{dA}\right|^2 + b^2A^2\left|\chi^\mathrm{RD}(k,A)\right|^2\right\}~~,
 \label{rd-8}
\end{eqnarray}
where $\Omega_\mathrm{R}$ denotes the present day dimensionless energy density of radiation, $g_{r,eq}$ and $g_{r,0}$ represent the number of relativistic degrees of freedom at matter-radiation equality and today respectively. Furthermore by using Eq.~(\ref{rd-3}), we determine different terms present in the expression of $\Omega_\mathrm{GW}^\mathrm{(0)}(k)$, in particular,
\begin{eqnarray}\label{chi-0}
    \left|\chi^\mathrm{RD}(k,A)\right|^2 = \frac{1}{b^2A^2}&\Bigg\{&bA_\mathrm{re}\chi^\mathrm{RH}(k,A_\mathrm{re})\cos{(bA-bA_\mathrm{re})}+\chi^\mathrm{RH}(k,A_\mathrm{re})\sin{(bA-bA_\mathrm{re})}\nonumber\\
    &+&A_\mathrm{re}\frac{d\chi^\mathrm{RH}(k,A_\mathrm{re})}{dA}\sin{(bA-bA_\mathrm{re})}\Bigg\}^2
\end{eqnarray}
and
\begin{eqnarray}\label{chi'_0}
    \left|\frac{d\chi^\mathrm{RD}(k,A)}{dA} \right|^2 = \frac{1}{b^2A^4}\Bigg\{&\Big(&\chi^\mathrm{RH}(k,A_\mathrm{re})+A_\mathrm{re}\frac{d\chi^\mathrm{RH}(k,A_\mathrm{re})}{dA}\Big)\Big[\sin{(bA-bA_\mathrm{re})}-bA\cos{(bA-bA_\mathrm{re})}\Big]\nonumber\\
    &+&b^2AA_\mathrm{re}\chi^\mathrm{RH}(k,A_\mathrm{re})\sin{(bA-bA_\mathrm{re})}\Bigg\}^2 \ ,
\end{eqnarray}
respectively. Thus as a whole, the dimensionless energy density of GWs at present epoch (symbolized by $\Omega_\mathrm{GW}^\mathrm{(0)}(k)$) is given by Eq.~(\ref{rd-8}) where the respective quantities are determined above. In general, $\Omega_\mathrm{GW}^\mathrm{(0)}(k)$ can be expressed by an amplitude and a spectral tilt (w.r.t. the wave number) which we will determine for the modes $k<k_\mathrm{f}$, in the following sections. For this purpose, we need to understand that the explicit $k$ dependency on $\Omega_\mathrm{GW}^\mathrm{(0)}(k)$ comes through $\chi^\mathrm{RH}(k,A_\mathrm{re})$ (and its derivative as well) which follows Eq.~(\ref{chi_reheating}). Here it may be noted that the transfer function during the reheating depends on the ratio $\frac{k}{k_\mathrm{re}}$ as well as on $\frac{k}{k_\mathrm{f}}$. Owing to this fact, we will individually determine $\Omega_\mathrm{GW}^\mathrm{(0)}(k)$ for the two cases given by: (I) $k<k_\mathrm{re}$ and (II) $k_\mathrm{re} < k < k_\mathrm{f}$ respectively.

\subsection*{$\boxed{\text{Case (I): $k < k_\mathrm{re}$}}$}

The range $k<k_\mathrm{re}$ represent the modes which re-enter the horizon during the radiation dominated era. Due to the condition $k < k_\mathrm{re}$, one may use the asymptotic expression for the Bessel function
\begin{equation}
    \lim_{x\ll 1} J_\nu(x) = \frac{x^\nu}{2^\nu\Gamma(\nu+1)}\ ,
    \nonumber
\end{equation}
for all the Bessel functions present in the solution of $\chi^\mathrm{RH}(k,A)$ (see Eq.~(\ref{chi_reheating})). By utilizing this asymptotic form, we get the functions $\mathscr{D}$ and $\mathscr{N}$ ( appearing in Eq.~(\ref{chi_reheating}) ) at $A=A_\mathrm{re}$ as follows:
\begin{equation}\label{CI--2}
    \mathscr{D}\big|_{A_\mathrm{re}}=-4k_\mathrm{f}(1+3w_\mathrm{eff})\frac{1}{\Gamma(1-\nu)\Gamma(\nu)}
\end{equation}
and
\begin{equation}\label{CI--1}
    \mathscr{N}\big|_{A_\mathrm{re}} = \mathscr{D}\big|_{A_\mathrm{re}} - \frac{2k_\mathrm{f}}{(1+3w_\mathrm{eff})\Gamma(\nu+1)\Gamma(2-\nu)}\left(\frac{k}{k_\mathrm{re}}\right)^2\left(\frac{1}{A_\mathrm{re}}\right)^\frac{5+3w_\mathrm{eff}}{2}
\end{equation}
respectively, where we may recall the factor $\nu$ from Eq.~(\ref{Reh-4}). The above expressions immediately yield the transfer function at $A=A_\mathrm{re}$ as,
\begin{align}\label{chi_end_re}
    \chi^\mathrm{RH}(k,A_\mathrm{re})=\frac{\mathscr{N}}{\mathscr{D}}\Bigg|_{A_\mathrm{re}}
    =1-\frac{2}{3}\left(\frac{k}{k_\mathrm{re}}\right)^2\frac{1}{(1-w_\mathrm{eff})(5+3w_\mathrm{eff})}\left(\frac{1}{A_\mathrm{re}}\right)^\frac{5+3w_\mathrm{eff}}{2}
    = 1 + \mathcal{O}\left(k/k_\mathrm{re}\right)^2\ .
\end{align}
Moreover Eq.~(\ref{Reh-7}), due to the asymptotic form aforementioned, leads to
\begin{equation}\label{chi'_end_re}
    A_\mathrm{re}\frac{d\chi^\mathrm{RH}(k,A)}{dA}\bigg|_{A_\mathrm{re}}
    = \left(\frac{k}{k_\mathrm{re}}\right)^2\frac{1}{(5+3w_\mathrm{eff})}\left(\frac{1}{A_\mathrm{re}}\right)^\frac{5+3w_\mathrm{eff}}{2}
    = \mathcal{O}\left(k/k_\mathrm{re}\right)^2 \ .
\end{equation}
The above two equations clearly depict that in the leading order of $\frac{k}{k_\mathrm{re}} < 1$, the transfer function at the end of the reheating becomes unity and its derivative vanishes. This indicates that the transfer function over the modes $k < k_\mathrm{re}$ remains almost constant (which is $\approx 1$) from the end of the inflation to the end of the reheating. This is however expected as the modes $k < k_\mathrm{re}$ lie in the super-Hubble regime, where the tensor perturbation gets frozen, during the reheating stage. Plugging the expressions of Eqs.~(\ref{chi_end_re}) and (\ref{chi'_end_re}) into Eq.~(\ref{chi-0}) lead to,
\begin{eqnarray}
 \left|\chi^\mathrm{RD}(k,A)\right|^2 = \frac{1}{b^2A^2}&\Bigg\{&bA_\mathrm{re}\left(1 + \mathcal{O}\left(k/k_\mathrm{re}\right)^2\right)\cos{(bA-bA_\mathrm{re})}+\left(1 + \mathcal{O}\left(k/k_\mathrm{re}\right)^2\right)\sin{(bA-bA_\mathrm{re})}\nonumber\\
    &+&\mathcal{O}\left(k/k_\mathrm{re}\right)^2\sin{(bA-bA_\mathrm{re})}\Bigg\}^2 \approx \frac{1}{b^2A^2}\sin^2{(bA)}~~.
    \label{CI-1}
\end{eqnarray}
In arriving at the final result of Eq.~(\ref{CI-1}), we have used $bA \gg 1$ (as we are evaluating at late times during radiation domination) and $bA_\mathrm{re} = \frac{k}{k_\mathrm{re}} < 1$. Similarly by plugging the expressions of Eqs.~(\ref{chi_end_re}) and (\ref{chi'_end_re}) into Eq.~(\ref{chi'_0}) and after a little bit of simplification, we obtain
\begin{eqnarray}
 \left|\frac{d\chi^\mathrm{RD}(k,A)}{dA} \right|^2 \approx \frac{1}{A^2}\left\{\left(\frac{1}{bA} + bA_\mathrm{re}\right)\sin{(bA)} - \cos{(bA)}\right\}^2 \approx \frac{1}{A^2}\cos^2{(bA)}~~.
 \label{CI-2}
\end{eqnarray}
The above two expressions, due to Eq.~(\ref{rd-8}), immediately provide the dimensionless energy density of GWs today over the modes $k<k_\mathrm{re}$ as,
\begin{equation}\label{CI-3}
    \boxed{\ \Omega_\mathrm{GW}^\mathrm{(0)}(k)h^2 \simeq \left(\frac{1}{6\pi^2}\right)\Omega_\mathrm{R}h^2\left(\frac{H_\mathrm{i}}{M_\mathrm{Pl}}\right)^2 \ , \ }
\end{equation}
where we have assumed that $g_{r,0} = g_{r,eq}$. Therefore the GWs spectrum today for the modes $k<k_\mathrm{re}$ seems to be scale invariant and having the amplitude shown in Eq.~(\ref{CI-3}).

\subsection*{$\boxed{\text{Case (II): $k_\mathrm{re} < k < k_\mathrm{f}$}}$}

Let us now turn to the domain $k_\mathrm{re} < k < k_\mathrm{f}$ representing the modes that re-enter the horizon during the reheating era. We will start by mentioning that $\mathscr{D}|_\mathrm{re}$ in the present case still follows Eq.~(\ref{CI--2}), as it solely depends on $k/k_\mathrm{f}$ which continues to be less than unity. On contrary, $\mathscr{N}|_\mathrm{re}$ over the modes $k_\mathrm{re} < k < k_\mathrm{f}$ gets different than the previous case of Eq.~(\ref{CI--1}) due to its dependency on both the ratios $\frac{k}{k_\mathrm{f}}$ and $\frac{k}{k_\mathrm{re}}$. Now the Bessel functions containing $k/k_\mathrm{f}$ in its argument may be expressed by the asymptotic form as,
\begin{equation}
    \lim_{x\ll 1} J_\nu(x) = \frac{x^\nu}{2^\nu\Gamma(\nu+1)}\nonumber
\end{equation}
while the Bessel functions having $k/k_\mathrm{re}$ can be approximated by,
\begin{equation}
    \lim_{x\gg 1 }J_\nu(x) = \sqrt{\frac{2}{\pi x}}\cos\left(x-\frac{\pi}{2}\left(\nu+\frac{1}{2}\right)\right) \simeq \sqrt{\frac{2}{\pi x}}\cos{x} \ .\nonumber
\end{equation}
On utilizing such asymptotic behaviours of respective Bessel functions, we determine
\begin{equation}\label{CII-1}
    \chi^\mathrm{RH}(k,A_\mathrm{re})=\left(\frac{\Gamma(1-\nu)}{\sqrt{\pi}}\right)\big(1+3w_\mathrm{eff}\big)^\frac{2}{1+3w_\mathrm{eff}}\left(\frac{k}{k_\mathrm{re}}\right)^{\frac{-2}{1+3w_\mathrm{eff}}}\cos{\left(\frac{2k/k_\mathrm{re}}{1+3w_\mathrm{eff}}\right)}
\end{equation}
and
\begin{equation}\label{CII-2}
    A_\mathrm{re}\frac{d\chi^\mathrm{RH}(k,A)}{dA}\bigg|_{A_\mathrm{re}}=-\left(\frac{\Gamma(1-\nu)}{\sqrt{\pi}}\right)\big(1+3w_\mathrm{eff}\big)^\frac{2}{1+3w_\mathrm{eff}}\left(\frac{k}{k_\mathrm{re}}\right)^{\frac{3w_\mathrm{eff}-1}{3w_\mathrm{eff}+1}}\sin{\left(\frac{2k/k_\mathrm{re}}{1+3w_\mathrm{eff}}\right)} \ .
\end{equation}
Eq.~(\ref{CII-2}) clearly demonstrates that the transfer function over the modes $k_\mathrm{re} < k < k_\mathrm{f}$ does not remain frozen during the reheating stage. This is also reflected from Eq.~(\ref{CII-1}) which indicates that $\chi^\mathrm{RH}(k,A)$ at the end of reheating is different than unity, i.e., different than that of at the end of the inflation. Using the above two expressions into Eq.~(\ref{chi-0}) yields,
\begin{equation}\label{CII-3}
    \left|\chi^\mathrm{RD}(k,A)\right|^2 \approx \frac{1}{b^2A^2}\big(1+3w_\mathrm{eff}\big)^{\frac{4}{1+3w_\mathrm{eff}}}\left(\frac{\Gamma(1-\nu)}{\sqrt{\pi}}\right)^2\left(\frac{k}{k_\mathrm{re}}\right)^{2\left(\frac{3w_\mathrm{eff}-1}{3w_\mathrm{eff}+1}\right)}\cos^2{\left[bA+\frac{2k/k_\mathrm{re}}{1+3w_\mathrm{eff}}\right]}~~,
\end{equation}
while by plugging the expressions of Eqs.~(\ref{CII-1}) and (\ref{CII-2}) into Eq.~(\ref{chi'_0}), we obtain
\begin{equation}\label{CII-4}
    \left|\frac{d\chi^\mathrm{RD}(k,A)}{dA} \right|^2 \approx \frac{1}{A^2}\big(1+3w_\mathrm{eff}\big)^\frac{4}{1+3w_\mathrm{eff}}\left(\frac{\Gamma(1-\nu)}{\sqrt{\pi}}\right)^2\left(\frac{k}{k_\mathrm{re}}\right)^{2\left(\frac{3w_\mathrm{eff}-1}{3w_\mathrm{eff}+1}\right)}\sin^2{\left[bA+\frac{2k/k_\mathrm{re}}{1+3w_\mathrm{eff}}\right]}~~.
\end{equation}
The above two expressions, owing to Eq.~(\ref{rd-8}), result to the dimensionless energy density of GWs today for the domain $k_\mathrm{re} < k < k_\mathrm{f}$ as follows:
\begin{equation}\label{CII-5}
    \boxed{\ \Omega_\mathrm{GW}^\mathrm{(0)}(k)h^2 \simeq \left(\frac{1}{6\pi^2}\right)\Omega_\mathrm{R}h^2\left(\frac{H_\mathrm{i}}{M_\mathrm{Pl}}\right)^2 \big(1+3w_\mathrm{eff}\big)^\frac{4}{1+3w_\mathrm{eff}}\left(\frac{\Gamma(1-\nu)}{\sqrt{\pi}}\right)^2\left(\frac{k}{k_\mathrm{re}}\right)^{2\left(\frac{3w_\mathrm{eff}-1}{3w_\mathrm{eff}+1}\right)} \ , \ }
\end{equation}
where, once again, we have assumed that $g_{r,0} = g_{r,eq}$. Eq.~(\ref{CII-5}) clearly depicts that the GWs spectrum today over $k_\mathrm{re} < k < k_\mathrm{f}$ gets a tilted nature w.r.t. the wave number and the amount of tilt depends on $w_\mathrm{eff}$. Here $w_\mathrm{eff}$ is the reheating EoS parameter which is related with the model parameter $m$ via $w_\mathrm{eff} = -1+2/(3m)$. Therefore the tilt of the GWs spectrum (let us symbolize it by $n_\mathrm{GW}$) from Eq.~(\ref{CII-5}) turns out to be,
\begin{eqnarray}
 n_\mathrm{GW} = 2\left(\frac{3w_\mathrm{eff}-1}{3w_\mathrm{eff}+1}\right) = \frac{2(1-2m)}{1-m}~~.
 \label{CII-6}
\end{eqnarray}
Notably, $n_\mathrm{GW}$ vanishes for $m=1/2$. Moreover the spectrum is blue for $m < \frac{1}{2}$, otherwise for $m > \frac{1}{2}$, it becomes red tilted.

As a whole, by comparing Eq.~(\ref{CI-3}) and Eq.~(\ref{CII-5}), we may argue that the GWs spectrum today is flat for the modes that re-enter the horizon during the radiation domination, while the spectrum has a tilt over the modes re-entering the horizon during the reheating era and the amount of the tilt is fixed by the parameter $m$. In particular,
\begin{align}
\mathrm{Tilt~of~the~GWs~spectra} =
\begin{cases}
  0 \hspace{3.25cm} \text{for} \hspace{0.5cm} k < k_\mathrm{re} & \\
    n_\mathrm{GW} = \frac{2(1-2m)}{1-m} \hspace{1.cm} \text{for} \hspace{0.5cm} k_\mathrm{re}<k<k_\mathrm{f}~.
\end{cases}
\label{CII-7}
\end{align}

 \begin{figure}[!h]
\begin{center}
\centering
\includegraphics[width=3.2in,height=2.2in]{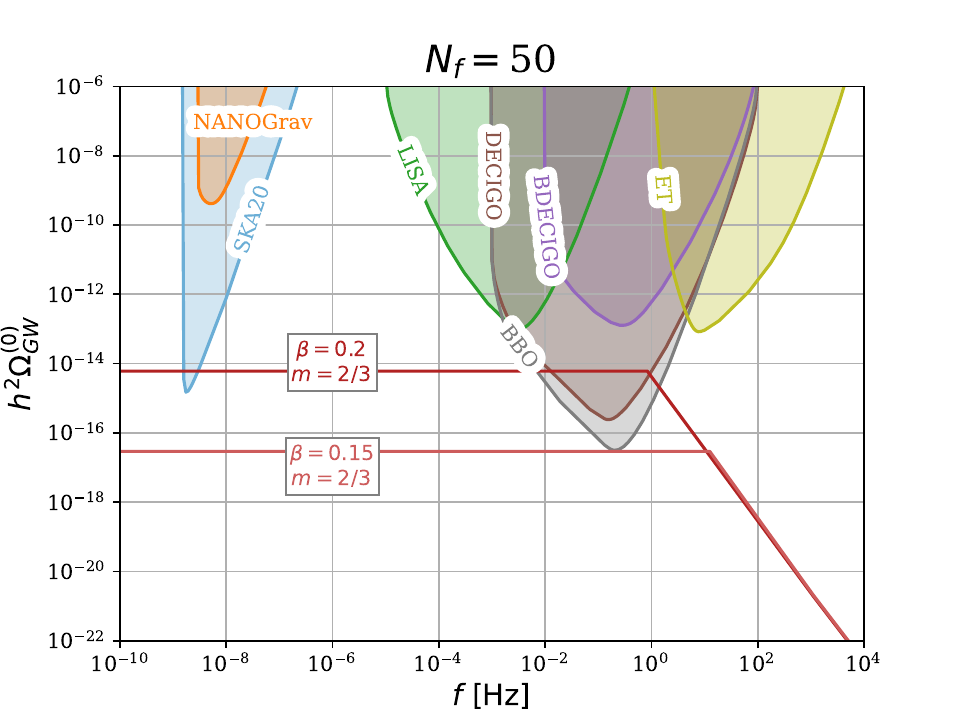}
\includegraphics[width=3.2in,height=2.2in]{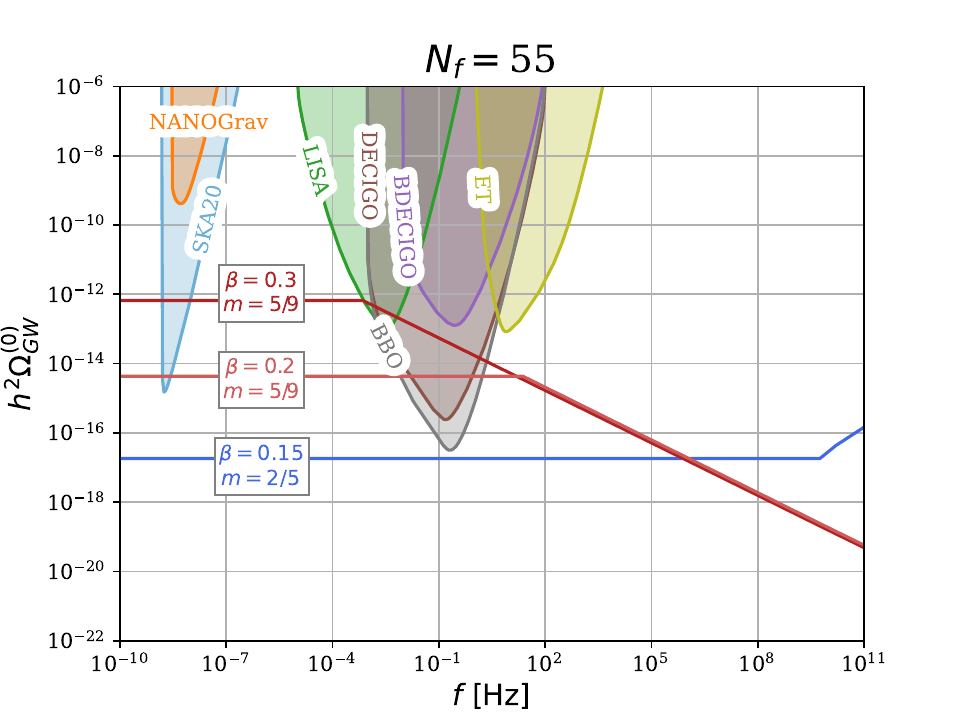}
\caption{{\color{blue}Left Plot}: $\Omega_\mathrm{GW}^\mathrm{(0)}$ vs. $\mathrm{f}$[Hz] for $N_\mathrm{f} = 50$; {\color{blue}Right Plot}: $\Omega_\mathrm{GW}^\mathrm{(0)}$ vs. $\mathrm{f}$[Hz] for $N_\mathrm{f} = 55$. In both the plots, we consider a set of values of the entropic parameters $\beta$ and $m$, and moreover, the other entropic parameters, namely $\sigma_0$ and $\alpha_{+}$, are taken as $\sigma_0= 0.015$ and $\alpha_{+}/\beta = 10^{-6}$. Such values of entropic parameters are indeed consistent with their viable ranges coming from the inflation and the reheating phenomenology, see Table.~[\ref{Table-0}]. Clearly the GWs spectra is flat for $k<k_\mathrm{re}$, while it has a non-zero tilt in the domain $k_\mathrm{re} < k < k_\mathrm{f}$. In particular, we take $m = 2/3$, $5/9$ and $2/5$ which lead to the indices $n_\mathrm{GW} = -2$, $-1/2$ and $2/3$ respectively, as expected from Eq.~(\ref{CII-7}). Furthermore, the constant amplitude of the GWs spectra associated with the modes $k < k_\mathrm{re}$ seems to decrease as the value of $\beta$ decreases.}
\label{plot-R1}
\end{center}
\end{figure}

 \begin{figure}[!h]
\begin{center}
\centering
\includegraphics[width=4.0in,height=3.0in]{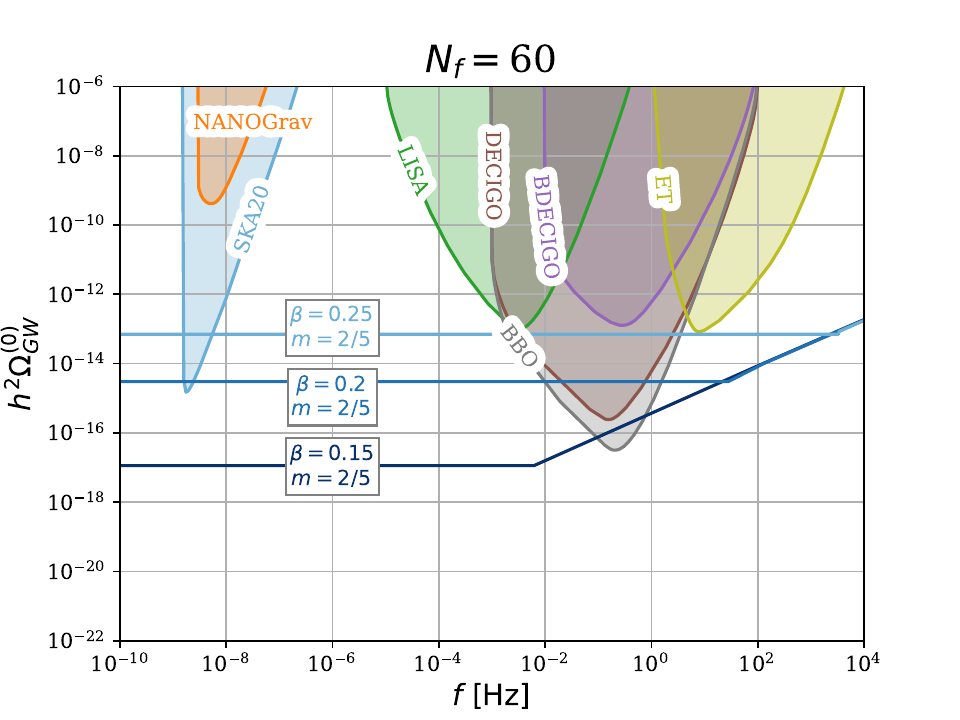}
\caption{$\Omega_\mathrm{GW}^\mathrm{(0)}$ vs. $\mathrm{f}$[Hz] for $N_\mathrm{f} = 60$. Similar to the previous figure, we consider a set of values of the entropic parameters $\beta$ and $m$, and the other entropic parameters, namely $\sigma_0$ and $\alpha_{+}$, are taken as $\sigma_0= 0.015$ and $\alpha_{+}/\beta = 10^{-6}$. Clearly the GWs spectra is flat for $k<k_\mathrm{re}$, while it has a non-zero tilt in the domain $k_\mathrm{re} < k < k_\mathrm{f}$. Here one may recall that for $N_\mathrm{f} = 60$ (in particular, for $N_\mathrm{f} \gtrsim 57.3$), the background phenomenology puts a constraint on $m < 1/2$ which in turn makes $n_\mathrm{GW} > 0$. Taking this into account, here we take $m = 2/5$ leading to $n_\mathrm{GW} = 2/3$, as expected from Eq.~(\ref{CII-7}). Furthermore, the constant amplitude of the GWs spectra associated with the modes $k < k_\mathrm{re}$ seems to decrease as the value of $\beta$ decreases.}
\label{plot-R2}
\end{center}
\end{figure}

In Fig.~[\ref{plot-R1}] and [\ref{plot-R2}], we have plotted the spectrum of GWs today that arise in the present context of horizon cosmology for a set of values of the entropic parameters $\beta$, $\sigma_0$ and $m$, and moreover, we have considered three different inflationary e-folding number: $N_\mathrm{f} = 50$, $55$ and $60$ respectively. The set of values of the entropic parameters are chosen from their viable ranges based on the Table.~[\ref{Table-0}]. The figures clearly illustrates the qualitative features as discussed above: (i) the spectrum is strictly scale invariant in the domain $k<k_\mathrm{re}$, and (ii) the spectrum has the tilt $n_\mathrm{GW}$ over the modes $k_\mathrm{re} < k < k_\mathrm{f}$. In order to realize such features directly from the figures, we should mention the dependency of $k_\mathrm{re}$ on the entropic parameters, and it goes as,
\begin{eqnarray}
 k_\mathrm{re} = a_\mathrm{re}H_\mathrm{re} = \left(a_\mathrm{i}H_\mathrm{i}\right)\left(\frac{H_\mathrm{f}}{H_\mathrm{i}}\right)\left(\frac{a_\mathrm{f}}{a_\mathrm{i}}\right)\left(\frac{a_\mathrm{re}}{a_\mathrm{f}}\right)\left(\frac{H_\mathrm{re}}{H_\mathrm{f}}\right) = k_\mathrm{CMB}\left(\frac{H_\mathrm{f}}{H_\mathrm{i}}\right)\mathrm{exp}\left[N_\mathrm{f} - \left(\frac{1-m}{m}\right)N_\mathrm{re}\right]~~,
 \label{CII-8}
\end{eqnarray}
where $k_\mathrm{CMB} = a_\mathrm{i}H_\mathrm{i}$ is the CMB scale mode ($\sim 0.05\mathrm{Mpc}^{-1}$) that crosses the horizon at the beginning of inflation, and $N_\mathrm{re}$ is the reheating e-fold number which, in terms of entropic parameters, is shown in Eq.~(\ref{et-10}). Coming to the figures, we have plotted the spectra for $m = 2/3$, $5/9$ and $2/5$ which lead to the indices $n_\mathrm{GW} = -2$, $-1/2$ and $2/3$ respectively, as expected from Eq.~(\ref{CII-7}). Here one may recall that for $N_\mathrm{f} = 60$ (particularly for $N_\mathrm{f} \gtrsim 57.3$), the background phenomenology puts a constraint on $m< 1/2$ which in turn makes the GWs spectra in the range $k_\mathrm{re} < k< k_\mathrm{f}$ blue tilted. Taking this into account, we consider $m=2/5$ for the set $N_\mathrm{f} = 60$, leading to $n_\mathrm{GW} = 2/3$. In the figures, we have also included the sensitivity curves of some of the current and forthcoming GW observatories. Interestingly, we find that for certain values of $\beta$ and $m$, the GWs spectra indeed intersect the sensitivity curves of some of the observatories like SKA, LISA, DECIGO, BBO etc. This in turn provides a possible way for measuring the generalized entropic parameters: $\beta$, $\sigma_0$ and $m$, respectively. In order to realize this, we need to understand the following points regarding the spectra: (i) the constant amplitude of $\Omega_\mathrm{GW}^\mathrm{(0)}(k)$, in the domain $k<k_\mathrm{re}$, depends on $\beta$ and $\sigma_0$, which is also evident from the figures since the amplitude seems to decrease as the value of $\beta$ decreases (and we consider $\sigma_0 = 0.015$ in all the plots); and (ii) $n_\mathrm{GW}$, i.e. the tilt of $\Omega_\mathrm{GW}^\mathrm{(0)}(k)$ associated with the modes $k_\mathrm{re} < k < k_\mathrm{f}$, depends solely on $m$ via $n_\mathrm{GW} = 2(1-2m)/(1-m)$. Therefore in the region $k_\mathrm{re} < k < k_\mathrm{f}$, the GWs spectrum today appears to be blue tilted for $m < 1/2$, while it is red tilted for $m > 1/2$. Such a GWs spectrum can reveal the entropic parameters by the following ways:
\begin{itemize}
 \item owing to the fact that $n_\mathrm{GW}$ depends only on $m$, the spectral tilt of $\Omega_\mathrm{GW}^\mathrm{(0)}(k)$ in the domain $k_\mathrm{re} < k < k_\mathrm{f}$ immediately depicts the value of one of the entropic parameters $m$;

 \item the other two parameters, namely $\beta$ and $\sigma_0$, can be determined from the constant amplitude of the spectrum within $k < k_\mathrm{re}$ and the location of $k_\mathrm{re}$ where the spectrum gets a characteristic change, respectively. Actually both of these quantities depend on ($\beta$, $\sigma_0$), and thus, by observing the constant amplitude of $\Omega_\mathrm{GW}^\mathrm{(0)}(k)$ in the domain $k<k_\mathrm{re}$ and identifying the location of $k_\mathrm{re}$, one may indirectly measure $\beta$ and $\sigma_0$.
\end{itemize}

In addition, we would like to mention that the theoretical expectation of GWs spectra in the present context of horizon cosmology does not intersect the sensitivity curve of the NANOGrav. This indicates that the standard inflationary evolution may not be the full story of early universe. Thus a modified inflationary evolution, for instance a short deceleration epoch inside inflation (and the deceleration epoch needs to be adjusted in such a way that the modes sensitive to the NANOGrav frequency cross the horizon during that deceleration era), may be required to corroborate the theoretical GWs spectra with the NANOGrav data.

\section{Conclusion}

In this work, we have attempted to constrain the entropic parameters corresponding to the 4-parameter generalized entropy from the possibility of primordial GWs spectrum in future observatories. The 4-parameter generalized entropy (symbolized by $S_\mathrm{g}$) is the minimal construction of generalized version of entropy that can reduce to all the known entropies proposed so far in the literature. Consequently it becomes utmost important to constrain the parameters present in $S_\mathrm{g}$, resulting to the motivation of the present work.

Regarding the background evolution, the 4-parameter generalized entropy successfully drives a viable and smooth evolution of the universe, started from inflation to reheating followed by a radiation era. In particular, the very early stage of the universe is described by a quasi de-Sitter inflation which has an exit at a finite e-fold number (around $N_\mathrm{f} = 50$ to $60$), and moreover, the observable indices (like the spectral index of primordial curvature perturbation and the tensor-to-scalar ratio) around the CMB scale are getting compatible with the recent Planck data for suitable ranges of the entropic parameters. After the inflation ends, the universe enters to a perturbative reheating stage when the entropic energy density decays to relativistic particles with a constant decay width. Such a reheating stage is generally parametrized by the respective e-fold number ($N_\mathrm{re}$) and the reheating temperature ($T_\mathrm{re}$), which eventually turns out to have a dependency on the entropic parameters. Here it deserves mentioning that the reheating era is not tightly constrained except the fact that the e-fold number of the reheating stage must be positive and the reheating temperature should be larger than the BBN temperature $\sim 10^{-2}\mathrm{GeV}$. Based on the conditions, $N_\mathrm{re} > 0$ and $T_\mathrm{re} > T_\mathrm{BBN}$, the entropic parameters get further constrained by the input of the reheating stage, see Table.~[\ref{Table-0}]. Thereafter in the perturbative level, we investigate whether the viable ranges of the entropic parameters (coming from the background level) allow the primordial GWs spectrum at present epoch to pass through the sensitivity curves of various GWs observatories. For this purpose, we determine the cosmic evolution of transfer function of the GWs spectrum generated during inflation in the present context of horizon cosmology, and in this regard, we confine ourselves to the modes $k < k_\mathrm{f}$ (where $k_\mathrm{f}$ represents the mode that crosses the horizon at the end of inflation). It turns out that the GWs spectrum today is flat for the modes that re-enter the horizon during the radiation era ($k<k_\mathrm{re}$), while the spectrum carries a tilted nature over the modes re-entering the horizon during the reheating stage ($k_\mathrm{re} < k< k_\mathrm{f}$). This is a direct consequence of the fact that the transfer function associated with the modes $k<k_\mathrm{re}$ lies in the super-Hubble scale and remains frozen during the reheating stage, unlike to the case of $k_\mathrm{re} < k< k_\mathrm{f}$ which already re-enters the horizon and the corresponding transfer function is not frozen during the reheating stage. As a result, the horizon cosmology corresponding to the 4-parameter generalized entropy indeed leads to an enhanced GWs spectrum that passes through the sensitivity regions of the observatories like LISA, DECIGO, BBO etc., see Fig.~[\ref{plot-R1}] (and Fig.~[\ref{plot-R2}]). This in turn provides a possible way for measuring the generalized entropic parameters: $\beta$, $\sigma_0$ and $m$, respectively. In order to realize this, we need to understand some important features of the GWs spectrum in the present context: (i) $\Omega_\mathrm{GW}^\mathrm{(0)}(k)$ has a constant amplitude in the domain $k<k_\mathrm{re}$ (with $k_\mathrm{re}$ being the mode that re-enters the horizon at the end of reheating), and moreover, the amplitude depends on $\beta$ and $\sigma_0$ through the inflationary energy scale; (ii) $\Omega_\mathrm{GW}^\mathrm{(0)}(k)$ shows a characteristic change at $k = k_\mathrm{re}$, in particular, $\Omega_\mathrm{GW}^\mathrm{(0)}(k)$ converts from a flat spectrum to a tilted one at $k=k_\mathrm{re}$, and the amount of tilt depends solely on $m$ via $n_\mathrm{GW} = 2(1-2m)/(1-m)$. Therefore in the region $k_\mathrm{re} < k < k_\mathrm{f}$, the GWs spectrum today appears to be blue tilted for $m < 1/2$, while it is red tilted for $m > 1/2$. Such a GWs spectrum can reveal the entropic parameters by the following ways: (i) owing to the fact that $n_\mathrm{GW}$ depends only on $m$, the spectral tilt of $\Omega_\mathrm{GW}^\mathrm{(0)}(k)$ in the domain $k_\mathrm{re} < k < k_\mathrm{f}$ immediately depicts the value of one of the entropic parameters $m$; (ii) the other two parameters, namely $\beta$ and $\sigma_0$, can be determined from the constant amplitude of the spectrum within $k < k_\mathrm{re}$ and the location of $k_\mathrm{re}$ where the spectrum gets a characteristic change, respectively. Actually both of these quantities depend on ($\beta$, $\sigma_0$), and thus, by observing the constant amplitude of $\Omega_\mathrm{GW}^\mathrm{(0)}(k)$ in the domain $k<k_\mathrm{re}$ and identifying the location of $k_\mathrm{re}$, one may indirectly measure $\beta$ and $\sigma_0$. Therefore we may argue that if the future observatories can detect the signal of primordial GWs, then our theoretical expectation carried in the present work may provide a possible tool for the measurement of the generalized entropic parameters.

Finally we would like to mention that primordial GWs maybe searched from modified gravity (see \cite{Odintsov:2021kup}). Hence, the determination of entropy parameters from that and comparison with primordial GWs in modified gravity may propose deep connection between entropy cosmology and modified gravity cosmology.

\section{Appendix}\label{sec-appendix}

\begin{figure}[!h]
\begin{center}
\centering
\includegraphics[width=3.0in,height=2.0in]{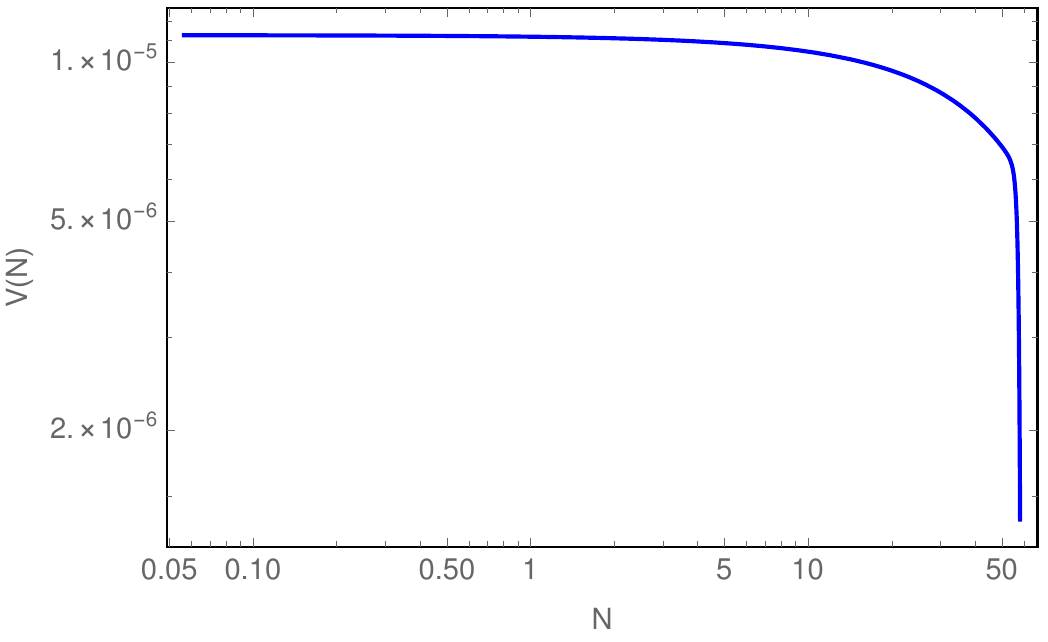}
\includegraphics[width=3.0in,height=2.0in]{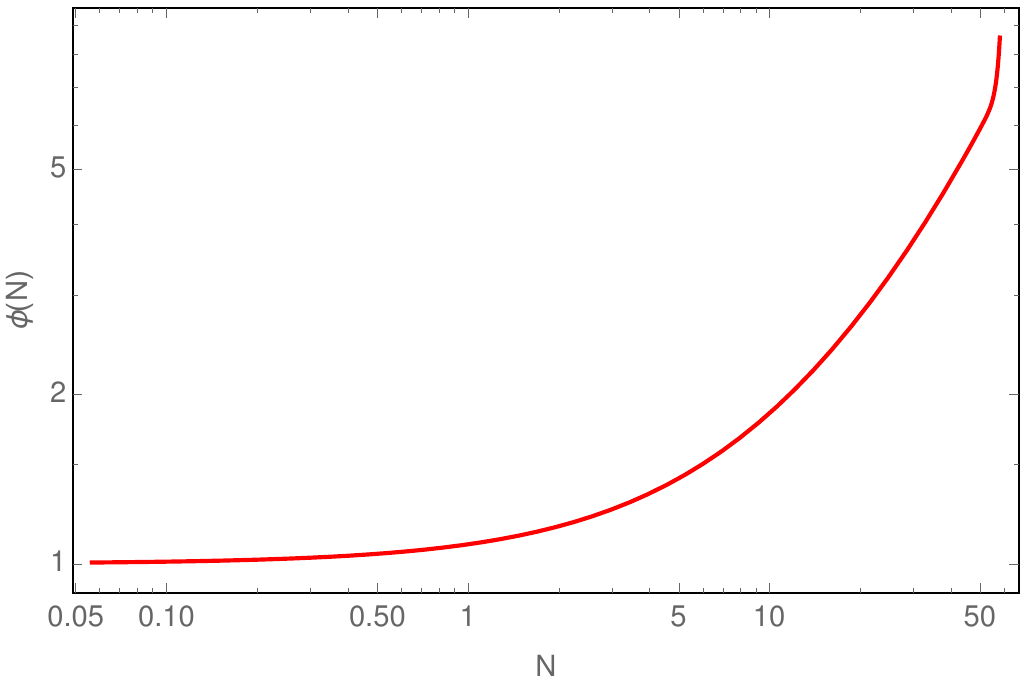}
\caption{{\color{blue}Left Plot}: $V(\phi(N))$ (in the unit of $M_\mathrm{Pl}^4$) vs. $N$; {\color{blue}Right Plot}: $\phi(N)$ (in the unit of $M_\mathrm{Pl}$) vs. $N$ for $N_\mathrm{f} = 58$. In both the plots, we consider $\sigma_0 = 0.015$ and $\beta=0.25$ which are well within the viable constraints coming from the inflationary phenomenology.}
\label{plot-1-app}
\end{center}
\end{figure}

In a canonical scalar-tensor (ST) theory, the Friedmann equations (for homogeneous, isotropic and spatially flat universe) read as,
\begin{eqnarray}
 3M_\mathrm{Pl}^2H^2&= &\frac{1}{2}\dot{\phi}^2 + V(\phi) \,,\nonumber\\
 2M_\mathrm{Pl}^2\dot{H}&=&-\dot{\phi}^2 \,,
 \label{app-1}
\end{eqnarray}
respectively, where $\phi$ is the scalar field (which is taken to be function of only the cosmic time due to the homogeneity of the spacetime) and $V(\phi)$ is its potential. Assuming the slow roll condition of the scalar field: $\dot{\phi}^2 \ll 2V(\phi)$, the above equations turn out to be,
\begin{eqnarray}
 3M_\mathrm{Pl}^2H^2&=&V(\phi) \,,\label{app-2a}\\
 2M_\mathrm{Pl}^2\dot{H}&=&-\dot{\phi}^2 \,.
 \label{app-2b}
\end{eqnarray}
With these set of equations, we now examine whether the solution of $H(N)$ obtained in the generalized entropic scenario can mimic a slow roll inflation under the canonical ST theory given by $\{\phi,V(\phi)\}$. Plugging $H(N)$ from Eq.~(\ref{solution-viable-inf-2}) to Eq.~(\ref{app-2a}), we reconstruct the form of $V(\phi(N))$ as follows:
\begin{eqnarray}
 V(\phi(N)) = 48\pi^2 M_\mathrm{Pl}^2\frac{\alpha_+}{\beta}
\left[\frac{2^{1/(\beta)}\exp{\left[-\frac{1}{\beta}\int^{N}\sigma(N)dN\right]}}
{\left\{1 + \sqrt{1 + 4\left(\alpha_+/\alpha_-\right)^{\beta}\exp{\left[-2\int^{N}\sigma(N)dN\right]}}\right\}^{1/(\beta)}}\right] \,.
\label{app-3}
\end{eqnarray}
To have a better understanding, we give the plot of $V(\phi(N))$ vs. $N$ for $N = [0,N_\mathrm{f}]$ (recall $N_\mathrm{f}$ is the total e-fold duration for the inflation in the entropic scenario), see the left plot of Fig.~[\ref{plot-1-app}]. The figure demonstrates that $V(\phi(N))$ remains almost constant except around $N \approx N_\mathrm{f}$ where the potential shows a rapid change, that pointing towards a slow roll inflation under the given ST theory. However the slow roll condition, i.e., $\dot{\phi}^2 \ll 2V(\phi)$ needs to be properly examined, which also requires the evolution of the scalar field. For this purpose, we use the solution of $H(N)$ in Eq.~(\ref{app-2b}) to get,
\begin{eqnarray}
 \frac{1}{M_\mathrm{Pl}}\frac{d\phi}{dN} = \left[\frac{2\sigma(N)}
{2\beta\sqrt{1 + 4\left(\alpha_+/\alpha_-\right)^{\beta}\exp{\left[-2\int_0^{N}\sigma(N)dN\right]}}}\right]^{1/2} \,,
\label{app-4}
\end{eqnarray}
(where we take only the positive mode of $d\phi/dN$). Owing to the complicated nature, we numerically solve Eq.~(\ref{app-4}) for $\phi = \phi(N)$ which is shown in the right plot of the Fig.~[\ref{plot-1-app}].

\begin{figure}[!h]
\begin{center}
\centering
\includegraphics[width=3.0in,height=2.0in]{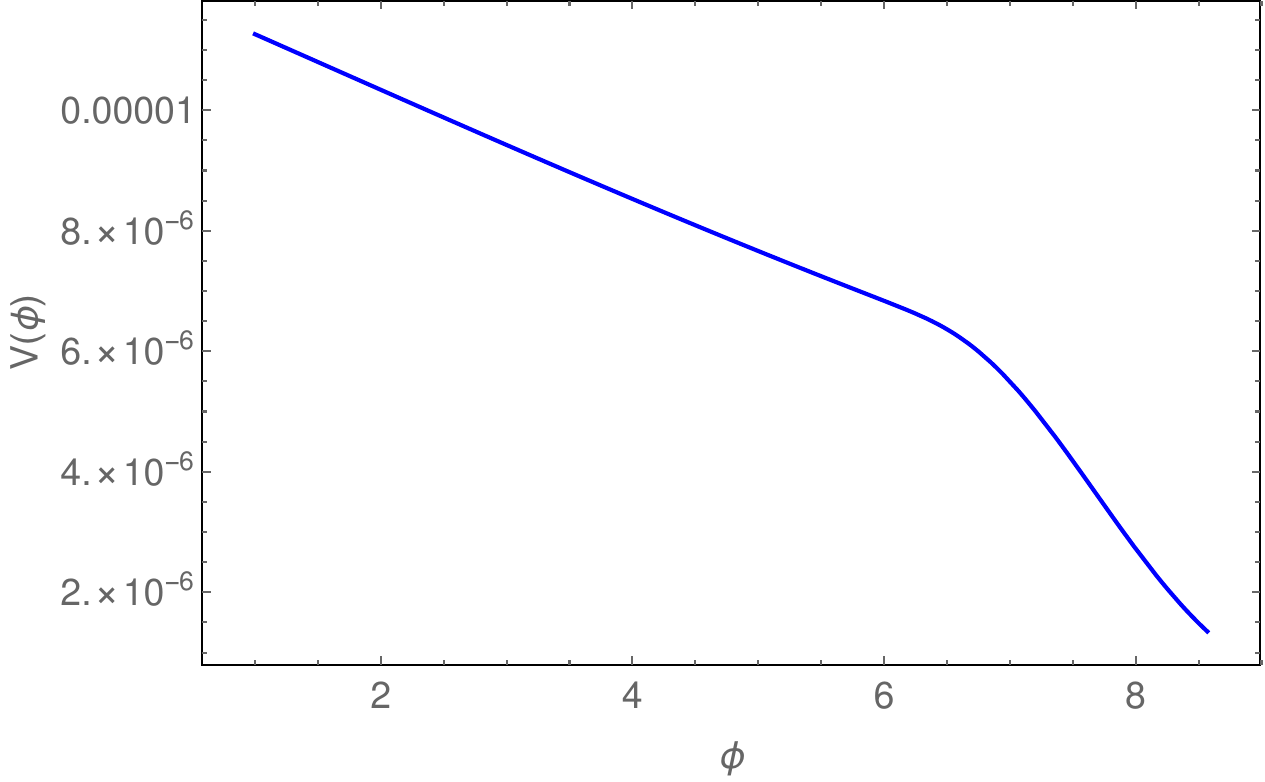}
\includegraphics[width=3.0in,height=2.0in]{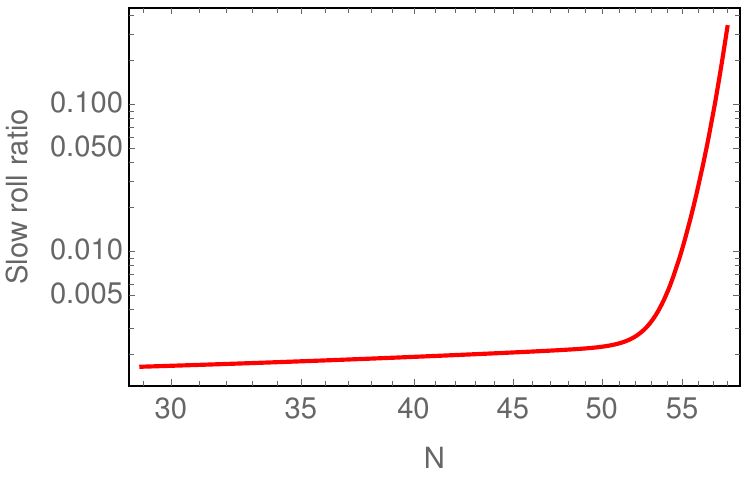}
\caption{{\color{blue}Left Plot}: $V(\phi)$ (in the unit of $M_\mathrm{Pl}^4$) vs. $\phi$ (in the unit of $M_\mathrm{Pl}$); {\color{blue}Right Plot}: $\dot{\phi}^2/(2V)$ vs. $N$ for $N_\mathrm{f} = 58$. In both the plots, we consider $\sigma_0 = 0.015$ and $\beta=0.25$ which are well within the viable constraints coming from the inflationary phenomenology.}
\label{plot-2-app}
\end{center}
\end{figure}

For completeness, we also present the variation of the scalar potential $V(\phi)$ with the scalar field $\phi$ by using $V = V(\phi(N))$ and $\phi = \phi(N)$ (we use the ``ParametricPlot'' in M{\small{ATHEMATICA}}), see the left plot of Fig.~[\ref{plot-2-app}]. Having obtained $V(\phi(N))$ and $\phi(N)$, we now explicitly examine the validity of the slow roll condition. In particular, by using Eq.~(\ref{app-3}) and Eq.~(\ref{app-4}), we give the plot of the ratio $\dot{\phi}^2/(2V)$ vs. $N$ in the right plot of the Fig.~[\ref{plot-2-app}] which indeed ensures the validation of the slow roll inflation under the canonical ST theory designated by $\{\phi,V(\phi)\}$.

Thus as a whole, the solution of $H(N)$ in the present context of generalized entropic cosmology can mimic a slow roll inflation under the canonical ST theory where $V(\phi)$ and $\phi(N)$ are given by Fig.~[\ref{plot-1-app}].

\end{document}